\documentclass[manuscript]{acmart}
\usepackage{mystyle}
\usepackage{booktabs}
\usepackage{graphicx}
\usepackage{lscape}

\usepackage{caption}
\AtBeginDocument{%
  \providecommand\BibTeX{{%
    \normalfont B\kern-0.5em{\scshape i\kern-0.25em b}\kern-0.8em\TeX}}}

\setcopyright{acmcopyright}
\copyrightyear{2024}
\acmYear{2024}
\acmDOI{XXXXXXX.XXXXXXX}

\acmConference[Conference acronym 'XX]{Make sure to enter the correct
  conference title from your rights confirmation email}{June 03--05,
  2018}{Woodstock, NY}
\begin{document}

\title[Improving Engagement and Efficacy of mHealth Micro-Interventions]{Improving Engagement and Efficacy of mHealth Micro-Interventions for Stress Coping: an In-The-Wild Study}

\author{Chaya Ben Yehuda}
\authornote{Both authors contributed equally to this research.}
\email{chayab1@mail.tau.ac.il}
\affiliation{%
  \institution{Tel Aviv University}
  \streetaddress{P.O. Box 39040}
  \city{Tel Aviv }
  \country{Israel}
  \postcode{6139001}
}

\author{Ran Gilad-Bachrach}
\affiliation{%
  \institution{Tel Aviv University}
  \streetaddress{P.O. Box 39040}
  \city{Tel Aviv}
  \country{Israel}}
\email{rgb@tauex.tau.ac.il}

\author{Yarin Udi}
\authornotemark[1]
\email{yarinudi@mail.tau.ac.il}
\affiliation{%
  \institution{Tel Aviv University}
  \streetaddress{P.O. Box 39040}
  \city{Tel Aviv }
  \country{Israel}
  \postcode{6139001}
}

\renewcommand{\shortauthors}{Ben-Yehuda, Gilad-Bachrach \& Udi}
\begin{abstract}
Sustaining long-term user engagement with mobile health (mHealth) interventions while preserving their high efficacy remains an ongoing challenge in real-world well-being applications. To address this issue, we introduce a new algorithm, the Personalized, Context-Aware Recommender (PCAR), for intervention selection and evaluate its performance in a field experiment.

In a four-week, in-the-wild experiment involving 29 parents of young children, we delivered personalized stress-reducing micro-interventions through a mobile chatbot. We assessed their impact on stress reduction using momentary stress level ecological momentary assessments (EMAs) before and after each intervention.

Our findings demonstrate the superiority of PCAR intervention selection in enhancing the engagement and efficacy of mHealth micro-interventions to stress coping compared to random intervention selection and a control group that did not receive any intervention. 
Furthermore, we show that even brief, one-minute interventions can significantly reduce perceived stress levels (p=0.001). 
We observe that individuals are most receptive to one-minute interventions during transitional periods between activities, such as transitioning from afternoon activities to bedtime routines. 

Our study contributes to the literature by introducing a personalized context-aware intervention selection algorithm that improves engagement and efficacy of mHealth interventions, identifying key timing for stress interventions, and offering insights into mechanisms to improve stress coping.

\end{abstract}

\ccsdesc[500]{Human-centered computing~Empirical studies in HCI}
\ccsdesc[300]{Applied computing~Health informatics}

\keywords{mobile micro-interventions, stress reduction, mental preparedness exercises, parental stress, Reinforcement Learning, Digital Health, mHealth, just-in-time}

\received{14 September 2023}
\maketitle
\section{Introduction}
\label{section: Introduction}
Stress is a pervasive issue with far-reaching implications for both physical and mental well-being~\cite{kalia2002assessing}. It is a complex psychological and physiological response to external or internal stimuli, commonly referred to as stressors~\cite{chrousos1992concepts}. The body's stress response is often characterized by a cascade of hormonal changes that affect various cognitive functions and overall health~\cite{mendl1999performing, arnsten2009stress}. The duration of a stressor is one of its most defining features and can be categorized into three timescales: acute stressors, daily events, and chronic stressors. By closely examining the response trajectory of a single acute stressor in real-time, one can study an individual's kinetics of stress response, including anticipation, peak reactivity, recovery, and regulation processes~\cite{epel2018more}.

Recent advancements in ubiquitous technology, such as the proliferation of smartphones and wearable devices, facilitate delivering timely and accessible mobile health (mHealth) interventions for stress reduction anytime and anywhere~\cite{silva2015mobile, prescott2022efficacy}. Despite the ease of delivery, users engagement with these interventions displays variability across different studies and generally proves lower in real-world applications~\cite{baumel2019there, baumel2019objective}. Particularly, sustaining long-term engagement remains an ongoing challenge. For example, \citet{paredes2014poptherapy} observed reducing engagement during a four-week naturalistic study due to repetitive interventions, leading to the introduction of the Last Switch Dependent (LSD) setting to vary intervention content~\cite{laforgue2022last}.
A comprehensive review by \citet{borghouts2021barriers} underscores the significance of contextual factors as important determinants influencing engagement rates, endorsing the use of machine learning algorithms to deliver context-aware, personalized interventions to boost user engagement~\cite{doherty2012engagement,liao2020personalized,konrad2015finding,schueller2010preferences}.

In this study, we introduce a novel algorithm for Personalized, Context-Aware intervention Recommendation (PCAR). This algorithm selects interventions by leveraging context variables, including personality traits and mobile sensor data. It also accounts for intervention fatigue, the tendency for diminishing returns from repeated use of the same intervention~\cite{paredes2014poptherapy,laforgue2022last}. Consequently, the algorithm aims to enhance the efficacy and long-term engagement with mobile health (mHealth) micro-interventions.

To investigate our algorithm's performance, we conducted a four-week, in-the-wild study with 29 parents of young children, for whom stress management is especially critical. 
We developed a system that enables real-time interactions with multiple participants simultaneously and delivered micro-interventions via a WhatsApp chatbot. Through the chatbot, we delivered ecological momentary assessments (EMAs)~\cite{shiffman2008ecological}, surveys, and intervention content. 

Our interactions were motivated by the emergence of Just-in-Time Adaptive Interventions (JITAIs) aimed at addressing the challenge of effective stress management~\cite{schroeder2020data,sano2017designing} via notifications on smartphones and wearables~\cite{baumel2020digital}. Existing JITAIs encounter limitations in two primary domains: content personalization and intervention timing~\cite{howe2022design}.
While technological advancements in JITAI frameworks have aimed to enhance engagement through timely intervention delivery~\cite{rozet2019using}, the primary focus has been on peak stress episodes~\cite{konrad2015finding,can2020relax, hovsepian2015cstress}. This narrow focus neglects the essential aspect of timing interventions across the complete stress response cycle, including the anticipation and recovery phases. 

The main objective of this work is to evaluate PCAR's ability to maintain efficacy over time of mHealth micro-interventions (RQ1), which also serves as an evaluation of the suitability of the LSD setting for addressing the information fatigue problem. 
We initiated both random and \pcar micro-interventions to examine their impacts on daily stress levels. Short-term stress reduction was assessed through momentary stress level ratings before and 10 minutes after each intervention via EMAs, while long-term effects were measured using pre-, mid-, and post-experiment Perceived Stress Scale questionnaire (PSS)~\cite{cohen1983global}. 

Our results demonstrate that even brief, one-minute interventions can significantly reduce perceived stress levels (p=0.001), regardless of the time of day. We also show the superiority of PCAR in intervention selection to enhance the engagement and efficacy of mHealth micro-interventions compared to random intervention selection.

Additionally, the study provides some information about how individuals engage with stress-reducing micro-interventions across different stress stages and time frames along the day (RQ2), as well as how they perceive and experience these interventions integrated into their daily lives (RQ3). We observe that individuals are most receptive during transitions between activities, serving as both "cool-down" and "warm-up" exercises for subsequent engagements. Hence, pinpointing the optimal timing (transitional periods) for these interventions is essential for maximizing engagement and improving the effectiveness of well-being applications. 

Our study contributes to the literature by introducing a personalized context-aware intervention selection algorithm that improves engagement and efficacy of mHealth interventions, identifying key timing for stress interventions, and offering insights into stress management and human-computer interaction.

The remainder of this paper is organized as follows: Section~\ref{section: Related Work} provides a literature review, Section~\ref{section:Methods} describes the methodology, Section~\ref{section: Finding} presents the results, Section~\ref{section: Limitations} presents limitation and Section~\ref{section: Conclusion} discusses the implications of our findings.

\section{Related Work}
\label{section: Related Work}
Stress is a pervasive issue that became even more common during and after the Covid-19 pandemic~\cite{american2019stress, jiang2020psychological,kalia2002assessing}. Stress responses are characterized by the disruption of the body's homeostasis in response to real or perceived threats or challenges (hereafter "stressors" or "stressor exposures")~\cite{chrousos1992concepts}, which can manifest in various detrimental effects, including cognitive impairments~\cite{mendl1999performing, arnsten2009stress}. 

\subsection{Technological Solutions for Stress Management}

Smartphone-based mental health applications have garnered considerable attention due to their ubiquity and seamless integration into everyday life~\cite{fuller2019randomized,linardon2019efficacy}. These consumer-oriented apps offer a diverse array of services, ranging from self-guided meditation platforms such as Headspace\footnote{\href{https://www.headspace.com/}{https://www.headspace.com/}} and Calm\footnote{\href{https://www.calm.com/}{https://www.calm.com/}}, to peer-support networks like Talklife\footnote{\href{https://www.talklife.com/}{https://www.talklife.com/}}, and online counseling services, for example, Talkspace\footnote{\href{https://www.talkspace.com/}{https://www.talkspace.com/}}. However, the majority of the existing stress management applications either necessitate user-initiated activation for stress mitigation activities or depend on pre-configured, context-agnostic notifications. These limitations highlight the absence of personalization, a crucial element for the effectiveness of stress management interventions~\cite{neary2018state,howe2022design}. 

Research on personalization in the context of stress management has primarily concentrated on two facets: delivery timing and content adaptation. The most advanced form of such personalization is known as Just-In-Time Adaptive Interventions (JITAI)~\cite{nahum2018just}. These interventions leverage ubiquitous sensing technologies to capture dynamic human behavior data, thereby delivering personalized, contextualized, and adaptable interventions. The objective is to improve efficacy by selecting the most appropriate treatment for a specific moment~\cite{nahum2018just,sano2017designing,smyth2016providing}. In the following sections, we will explore the recent advancements as well as the limitations associated with Just-in-time and its adaptive intervention components.

\subsubsection{Just-In-Time Technologies}
Recent technological advancements have considerably enhanced the capabilities of wearable biosensors, facilitating the collection of a wide range of physiological signals. This technological progress serves as a foundational element for the creation of personalized and real-time interventions in stress management~\cite{tazarv2021personalized, sarker2016finding, battalio2021sense2stop, burghardt2021having}. These interventions benefit from the real-time collection of diverse data points, such as busyness, geographical location, weather conditions, step count, and heart rate from users~\cite{battalio2021sense2stop,rehg2017mobile,sarker2016finding,morshed2022advancing}. Operating passively, these sensors continuously monitor individual stress levels, thereby identifying optimal moments for intervention deployment. Simultaneously, the domain has seen significant advancements in binary stress detection algorithms and pattern mining techniques, which further enhance the utility of sensor data in stress management frameworks~\cite{zhang2022wearable}.

The data patterns identified by these sensors, often preceding stress episodes, provide the ability to not only predict but also preemptively reduce stress. This shift from a largely reactive to a proactive approach has transformative implications for the field of stress management~\cite{kramer2019tomorrow}. Wearable devices like smartwatches and photo-plethysmography (PPG) sensors are suitable for extracting crucial physiological metrics such as heart rate (HR), heart rate variability (HRV), and skin conductance levels. However, the financial burden associated with these wearable devices often serves as an impediment to their broader adoption, thereby limiting the scalability of these devices~\cite{huarng2022adoption}. This highlights the pressing need for a framework that can achieve similar results using only the sensors integrated into mobile devices, thus democratizing access to effective stress management solutions.

While JITAI offers considerable promise, empirical studies have primarily focused on capturing moments of high stress~\cite{konrad2015finding,can2020relax}. Moreover, the implementation and efficacy of these interventions at scale in real-world settings remain in their initial stages. This gap in the existing methodologies accentuates the need for a more comprehensive approach that can capture not just the onset but also the various subsequent stages of stress as they manifest in real-world conditions.

\subsubsection{Adaptive Interventions Technologies}

The concept of personalization extends beyond the timing of interventions to include content that is either tailored for personal needs or selected to match personality and other context factors. Factors such as personality traits, demographic characteristics, and past experiences are instrumental in shaping an individual's vulnerability to stressors and their perception of stress intensity~\cite{bonanno2013regulatory,sur2014extending}. 

Research has demonstrated that participants who received interventions personalized to these individual characteristics exhibited a marked improvement in their coping skills compared to those who received non-specific, random interventions~\cite{paredes2014poptherapy, mishra2021detecting}. This underscores the importance of adaptive interventions that are not only contextually relevant but also intrinsically aligned with the individual's unique profile for more effective stress management.

One challenge encountered in previous implementations is the tendency of recommendation algorithms to generate repetitive recommendations, leading to user disengagement and boredom~\cite{paredes2014poptherapy, liao2020personalized, park2023understanding}. Addressing this issue without resorting to computationally intensive algorithms, such as full-fledged reinforcement learning approaches that necessitate large datasets and exhibit slow convergence rates~\cite{geva1993cartpole}, remains a significant obstacle.

The Last Switch Dependent (LSD) model offers a viable solution to this challenge by incorporating dynamic personalization. It monitors the "state" of each action (intervention), denoted by \(\tau\), to gauge the time elapsed since that particular action was last selected. Positive values of \(\tau\) signify that the action has been dormant for some time, while negative values indicate frequent recent selection. This enables the model to adapt its recommendations in real-time based on the user's most recent behavior. However, finding the optimal policy in the LSD setting is NP-hard, and the proposed approximation algorithm requires solving an integer linear programming problem in every round~\cite{laforgue2022last}. The LSD setting involves both a learning problem, that is, studying the expected reward of different actions given different states, and a planning problem of finding an optimal policy given these expected rewards.
\citet{papadigenopoulos2022non} and \citet{foussoul2023last} introduced improved approximation algorithm for the LSD planning problem with some additional constrains on the reward function. Moreover, while \citet{laforgue2022last} introduces the LSD setting and provides a theoretical analysis of the model, it was not tested in real-life intervention delivery settings. 

In this paper, we propose an economical algorithm for the LSD setting and evaluate it in the wild. More information can be found in Section~\ref{subsubsection: \pcar}.

\subsection{The Role of Timing in Coping Mechanisms and Interventions}
One of the most defining characteristics of a stressor is duration, which can be described in three timescales: acute stressors, daily events, and chronic stressors. Examining the response trajectory of one acute stressor using a magnifying glass in real time over minutes allows examination of an individual’s stress response kinetics, anticipation, peak reactivity, recovery, and regulation processes. The \textbf{Anticipation} stage is marked by pre-stressor rumination and elevated physiological arousal~\cite{niedenthal2005embodiment,van2017anticipatory}. At the \textbf{Peak} juncture, the body's stress response mechanisms are fully activated. Subsequently, in the \textbf{Recovery} stage, the body returns to its baseline physiological condition.
Each stage presents its unique challenges and opportunities for coping~\cite{epel2018more}.

Healthy coping mechanisms can reduce the negative impact of stress, while maladaptive strategies can exacerbate it, prolong each stage, and contribute to adverse long-term health outcomes~\cite{epel2018more}. 
In the \textbf{Anticipation} stage, emotional regulation strategies such as \textit{attentional deployment} are crucial for managing pre-stressor ruminations and physiological arousal~\cite{cohen2016stage,koole2011self,gross2015emotion}. Without effective coping, maladaptive strategies like excessive worrying can exacerbate stress at this stage. During the \textbf{Peak} stage, adaptive coping strategies like \textit{cognitive change} are particularly effective for managing the fully activated stress response~\cite{koole2011self,gross2015emotion}. In the \textbf{Recovery} stage, \textit{response modulation} strategies, including relaxation techniques and positive reframing, are essential for expediting the body's return to its baseline physiological state~\cite{koole2011self,gross2015emotion}. Maladaptive coping strategies, such as substance abuse, can delay recovery and contribute to sustained stress and associated health risks.

Specific interventions are effective at different stress stages. In the \textbf{Anticipation} stage, mindfulness, yoga~\cite{toussaint2021effectiveness,john2011effect}, visualization~\cite{felix2018guided},rhythmic activities~\cite{cairney2014uses}, and to-do lists ~\cite{16_Effective_Stress_Management}, have been identified as effective mechanisms for reducing anticipatory stress~\cite{vadoa1997relationship,koole2011self}. In high-performance domains, such as actors and athletes, these exercises not only mitigate anticipatory stress but also strongly correlate with their success~\cite{gould2009mental,van2019effects,filaire2007motivation,arvinen2007elite,taylor2008performance,zaza1994based}. During the \textbf{Peak} stage, cognitive-behavioral interventions (CBT) and Distracting activities shown to be effective~\cite{gross1998antecedent,sheppes2007better,koole2011self,nakao2021cognitive}. In the \textbf{Recovery} stage, relaxation techniques and scribbling aid in returning to baseline~\cite{Healthy_Coping_Skills,toussaint2021effectiveness,simmons2016keep,gilad2017scribbling}. Within these stages, special interest is directed towards the "about-to-happen" temporal window as a key moment for preemptive interventions for predictable stressors like public speaking~\cite{bassett1987physiological}, school tests~\cite{putwain2008examination}, or bedtime routines with young children~\cite{de2018parental}. Various studies corroborate this timing concept~\cite{rahman2016predicting,battalio2021sense2stop,sarker2016finding,mark2016email,howe2022design} suggesting that proactive approaches could significantly enhance intervention outcomes~\cite{tazarv2021personalized, sarker2016finding}. Complementing these findings, research by Gross underscores the importance of timing in emotional regulation strategies~\cite{gross1998antecedent}. \citet{sheppes2007better} suggests that intervening before the onset of stress can prevent the activation of the stress response, leading to more favorable outcomes in both acute and chronic stress scenarios. This preemptive approach is especially valuable in situations where stress is predictable, such as public speaking or challenges related to parenting such as morning and bedtime routines. Thus, the efficacy of various interventions is tied to the timing of their application.

Understanding the timing and peaks of stress is undeniably vital; however, theoretical knowledge about when to intervene constitutes only half of the equation. The other half resides in the efficacious delivery and sustained engagement with these interventions, particularly in real-world contexts where stressors are less predictable and more varied. Without robust engagement, even the most well-timed interventions risk becoming ineffectual. Therefore, attention is increasingly being channeled towards technological solutions that facilitate not only timely delivery but also foster effective engagement in stress management strategies~\cite{nahum2018just,howe2022design,doherty2012engagement}. 

\subsection{Stress Management in Parents of Young Children}
\label{subsection: Stress Management in Parents of Young Children}

Our research is specifically focused on parents of young children, a demographic for which three primary transitional periods have been identified as particularly stress-inducing: the transition from morning routines to work, the shift from work to afternoon activities, and the transition into bedtime routines~\cite{schneider2004stress,de2018parental}. This section offers a literature review centered on stress management strategies pertinent to this demographic.

Parenting stress is characterized by psychologists as the emotional tension experienced when one feels inundated by the responsibilities of parenthood, often attributed to a perceived insufficiency of resources to meet these demands~\cite{deater1998parenting, holly2019evidence}. Amidst the daily balancing act of managing multiple child-care tasks, parents frequently encounter stressful situations. Notably, parental stress levels have surged during the COVID-19 pandemic and have yet to revert to pre-pandemic norms~\cite{adams2021parents}, underscoring the necessity for augmented mental health resources and support mechanisms.

Three primary transitional periods have been pinpointed as especially stress-inducing: the transition from morning routines to occupational activities, the shift from work to afternoon engagements, and the transition into bedtime routines~\cite{schneider2004stress,de2018parental}. The morning separation from children is particularly highlighted as a potential stressor~\cite{drugli2012partnership}. A survey conducted by OnePoll, on behalf of Amazon Devices, polled 2,000 parents to ascertain the most stressful segments of their day. A majority of respondents identified morning preparations as the principal source of daily anxiety\footnote{\href{https://www.independent.co.uk/life-style/parents-mornings-stressful-childcare-poll-b1907699.html/}{https://www.independent.co.uk/life-style/parents-mornings-stressful-childcare-poll}}.

Comprehending these stress patterns is imperative for ascertaining the optimal timing for interventions. Such understanding also facilitates a nuanced analysis of the efficacy of interventions deployed before, during, and subsequent to these identified stress episodes.

\section{Methods}
\label{section:Methods}
\begin{figure}[tb]
    \centering
    \includegraphics[width=0.75\textwidth]{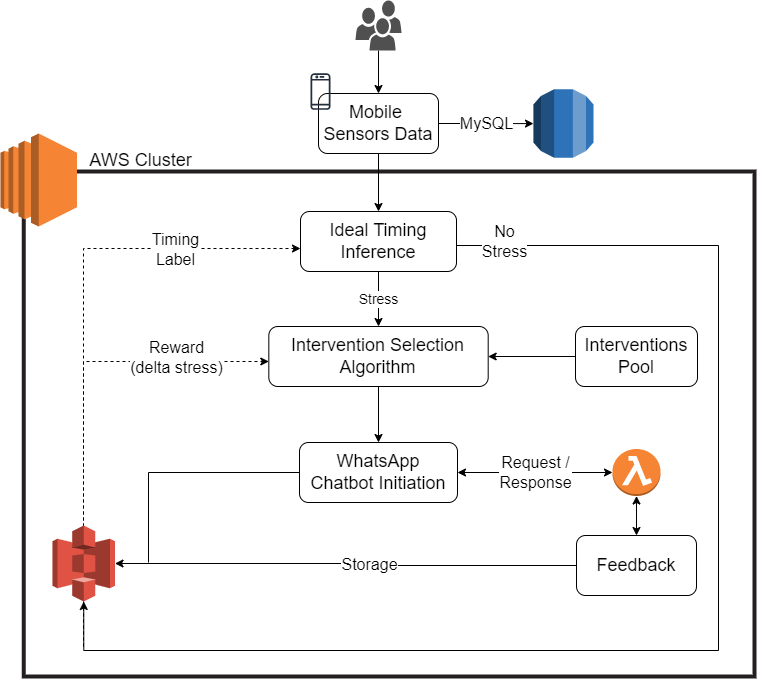}
    \caption{\textbf{The framework developed to improve engagement with the stress-reducing micro-interventions:} Mobile sensor data from the user is gathered via a specialized application and uploaded to a centralized database. A dedicated process analyzes this data to infer the most opportune moment for intervention, aiming to maximize user engagement. When the timing is deemed appropriate, a specific intervention is chosen, triggering a chatbot interaction with the user. User feedback is subsequently stored in the cloud for future analysis and refinement.}
    \label{fig:rycs_system.png}
\end{figure}

The primary objective of this research is to explore the short- and long-term effects of mHealth micro-interventions on daily life stress coping, and how personalization influences these effects, focusing on parents of young children, utilizing a comprehensive framework we have developed for this purpose (Figure~\ref{fig:rycs_system.png}). 
We also aim to enhance user engagement through personalized, context-aware recommendation (\pcarNS)~\cite{nahum2018just,christensen2009adherence,paredes2014poptherapy}.

The study employed a four-week, real-time, in-the-wild design with 29 parents, focusing on parents of young children. We used WhatsApp as the intervention delivery method, given its familiarity and popularity messaging platform, to deliver timely and accessible interventions. We compare \pcar to a random-based intervention recommender and a control one. The experiment was approved by institutional ethics committee.

\subsection{Personalized Context Aware Recommender (\pcarNS)}
\label{subsubsection: \pcar}
Here, we introduce the PCAR algorithm for personalizing intervention selection which dynamically tailors interventions to the specific needs and contextual factors of each participant.

The primary objective is to identify the most effective intervention for reducing stress levels. While previous research has modeled this problem as a multi-armed bandit issue, it was observed that such models tend to converge to selecting the same interventions repeatedly, leading to diminishing performance~\cite{paredes2014poptherapy}. Following~\citet{laforgue2022last} we argue for the necessity of a more sophisticated reinforcement learning model, yet, due to scarcity of feedback, since studies run for short periods with a limited number of participants, the model has to have limited capacity. Therefore, we adopt the Last Switch Dependent model~\cite{laforgue2022last}. In this setting, it is assumed that the reward of using an intervention depends only on the time since this arm (intervention) was switched in or out denoted by $\tau_a$ for the arm $a$. If $a$ was used consistently during the last $r$ rounds then $\tau_a = -r$ whereas if $a$ was not used in the last $r$ rounds then $\tau_a = r$.

In this setting, the state $\tau$ is fully observed. Moreover, the transition between states is known, however, the reward function itself is unknown and has to be learned. Towards this goal we introduce \pcar, a variant of the SARSA($\lambda$) algorithm~\cite{sutton2018reinforcement}. Unlike traditional SARSA, \pcar uses the structure of the Last Switch Dependent model to evaluate not just the current trajectory but also all potential trajectories that lead to the current state. This accelerates the learning process since every feedback received from a participant is used to update the $Q$ function multiple times. \pcar assumes that there are environmental factors (context) that are independent of the current state and encodes them in the policy as well. The details of \pcar are presented in Algorithm~1.

\begin{figure}[tbh]
    \centering
    \includegraphics[height=80mm, trim=125 420 100 100,clip]{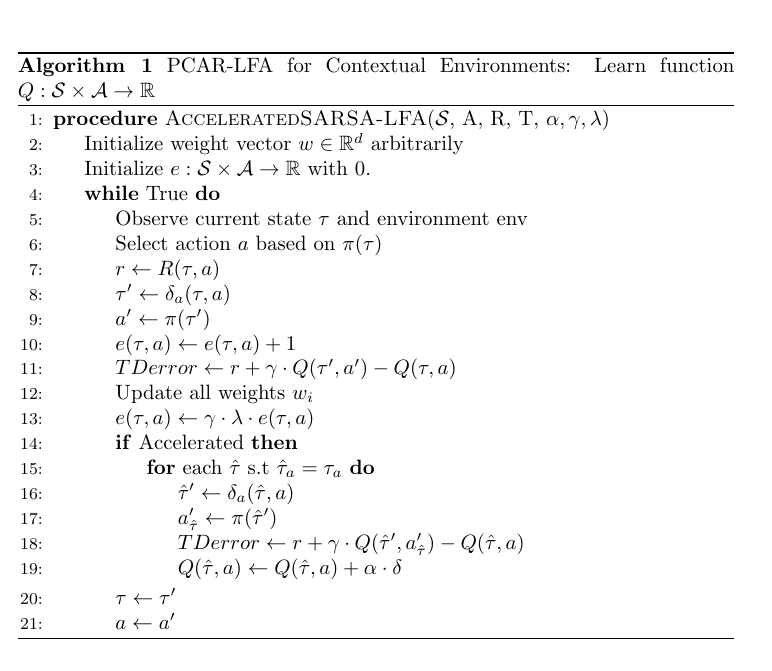}
    \label{fig:pcar_algo1.png}
\end{figure}

We define the state \( s \) as comprising two components: \( \tau \) and \emph{env}. Here, \( \tau \) captures the historical sequence of actions undertaken by the agent, akin to the Last Switch Dependent model \( \tau \). The \emph{env} component represents environmental factors independent of the agent's actions. A ghost state \( \hat{\tau} \) is defined for \( \tau \) if the expected rewards for actions are identical for both \( \tau \) and \( \hat{\tau} \), implying that their corresponding actions \( \hat{a}^{(i)} \) and \( a^{(i)} \) are identical. Using a known transition function \( \delta_a \)~\cite{laforgue2022last}, we then determine the subsequent ghost state \( \hat{\tau}_{a(t+1)} \) and the next ghost action \( \hat{a}^{(i)} \), updating the ghost state's weight as if it had been encountered.
Despite its prowess, \pcar grapples with computational and sample complexities. To elucidate, computational complexity concerns the resources (time and memory) an algorithm demands, whereas sample complexity relates to the interactions needed for satisfactory performance. In its raw form, \pcar demands exhaustive traversal of the state-action space, resulting in an impractical computational complexity of \(O(k^h)\) and high sample complexity.

The TensorAgent method deconstructs each action into an attribute dictionary, streamlining both complexities. An action \(a\) transforms from a solitary intervention to an attribute vector: 
\[ a = [a_1, a_2, \ldots, a_P] \] where each $a_i$ corresponds to an action attribute's possible value and $P$ represents the number of attributes. 
We ran independent copies of PCAR for selecting appropriate values for each attribute, we call these copies agents. Every copy (agent) received the same reward but had different actions to select from and hence leading to independent models.

As we bound the number of possible values an attribute can take by a constant $C$, the complexity becomes 
$O\left((P \cdot \tilde{C})^h\right)$. Given that $P \ll k$, it offers a significant reduction in complexity.
In practical terms, interventions like yoga are decomposed to attributes such as {'emotional regulation type': 'response modulation', 'therapy group': 'somatic', 'location': 'indoor'}.

\subsection{System Design}
\label{subsection: System Design}
\begin{figure}[tbh]
    \centering
    \includegraphics[width=0.8\textwidth]{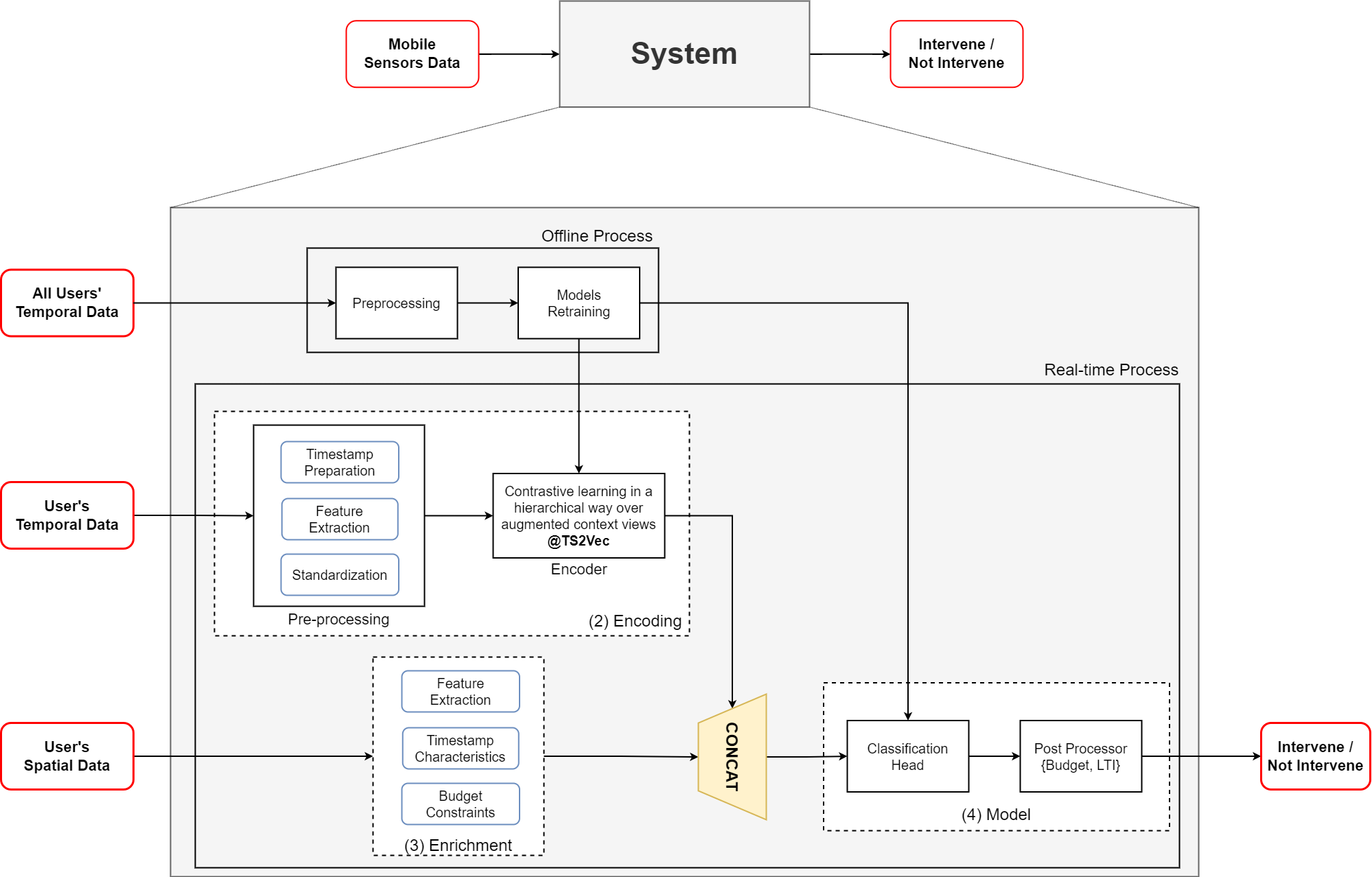}
    \caption{\textbf{Just-A-Minute Component:} Timing Inference Pipeline. The pipeline is divided into two main processes: the Offline Process, which involves dual jobs for model retraining using aggregated user data, and the Real-Time Process, which ingests and encodes data. A decision to initiate a stress intervention is made based on the calculated likelihood of user acceptance and current budget constraints.}

    \label{fig:detection_pipeline.png}
\end{figure}

To address our research aims, we have developed a system that enables real-time interactions with multiple participants simultaneously. The system can run autonomously 24/7 and utilizes sensor data from mobile devices, along with contextual features, to infer the appropriate time to initiate stress coping micro-interventions. The system comprises four primary components, as depicted in Figure~\ref{fig:detection_pipeline.png}. In the following sections, we provide detailed descriptions of each component:

\begin{enumerate}
    \item \textbf{Data Collection Component:} Tasked with gathering sensor data from participants' mobile devices via a dedicated application. To reduce power consumption on the phone, data was uploaded at approximately fifteen-minute intervals to our database for subsequent use in the timing detection algorithm and for offline analysis.
    
    \item \textbf{Just-A-Minute Component:} Utilizes the sensor data from participants' mobile devices to infer appropriate intervention timings. It aims to maximize the likelihood of user acceptance of interventions while adhering to constraints such as a maximum of three interventions per day and a minimum of two hours between consecutive interventions. More information can be found in Section ~\ref{subsubsection: Just-A-Minute Component}.
    
    \item \textbf{Personalized, Context-Aware Recommender Component:} Customizes the selection of interventions for each participant, taking into account factors like personality traits, current location, and past responses. This component is designed to mitigate the issue of repetitive recommendations, thereby reducing user disengagement~\cite{paredes2014poptherapy,liao2020personalized}. It is based on the implementation of our algorithm PCAR described in Section~\ref{subsubsection: \pcar}.
    
    \item \textbf{User Interface (UI) Component:} Comprises a WhatsApp chatbot engineered to enable user-friendly interactions. This serves as the main conduit for human-computer communication within our system.
\end{enumerate}

\subsubsection{Data Collection}
\label{subsubsection: data collection}
The primary input to our system consists of raw sensor data collected from mobile devices, including accelerometers, gyroscopes, and location sensors (a detailed list of sensors is provided in Appendix~\ref{appendix: sensors list}). These sensors were sampled continuously and concurrently using the \textit{AWARE-Light} application~\cite{ferreira2015aware}, which was specifically configured for this research. We selected this application after evaluating various other platforms, such as the \textit{mCerebrum application}~\cite{hossain2017mcerebrum} and the \textit{LAMP-platform}~\cite{rebecca_bilden_2023_7643628}. \textit{AWARE-Light} allowed for simultaneous data collection from an unlimited number of participants, offered customizable sampling configurations, and supports low sampling rates. However, the use of this application restricted our study to Android phone users exclusively.

Data was transmitted and uploaded to our databases at intervals ranging from 10 to 15 minutes, depending on the brand of the mobile device. We classified raw sensor data into two categories, which were subsequently utilized by the \textit{Just-A-Minute} inference component (Figure~\ref{subsubsection: Just-A-Minute Component}):

\begin{itemize}
    \item Spatial Sensor Data: These are static and non-motion sensors that provide insights into the device's current state or environmental conditions. Sensors in this category offer a snapshot of the present situation without necessarily monitoring time-based changes. They include the location, battery, Bluetooth, proximity, screen, and light sensors.
    \item Temporal Sensor Data: These sensors are dynamic and track changes over time. They capture motion, orientation, or other time-sensitive phenomena. The sensors in this category include the linear accelerometer, gyroscope, rotation, gravity, and magnetometer sensors. 
\end{itemize}

\subsubsection{Just-A-Minute Component}
\label{subsubsection: Just-A-Minute Component} This component was designed to infer the time for interventions for every active participant in the study independently. It adheres to a daily limit of three interventions, mandates a two-hour gap between each, and restricts delivery to between 8 AM and 9 PM. Two models are employed for this purpose: a neural network-based encoder and a classification head.

\begin{itemize}
    \item Encoder: Utilizes a \textit{TS2Vec} model~\cite{yue2022ts2vec} to learn a representation space that better describes the raw temporal sensor data for timing inference, this model harnesses data collected from all the participants. It employs hierarchical contrastive learning over augmented context views, enabling robust contextual representation for each timestamp.
    \item Classification Head: This part of the component uses both temporal and spatial raw data from a specific participant to infer the likelihood of intervention acceptance. It also leverages contextual features to adjust the prediction for different types of participants.
\end{itemize}

These models operate in two distinct modes — offline and real-time. The offline mode is scheduled to execute nightly at 2 AM, while the real-time mode runs at 5-minute intervals throughout the day. The offline process is tasked with training our models using the past day's streaming data, sensor readings, and participant feedback (binary label indicating whether the intervention's timing was good or not) from all active study participants. While, the real-time process utilizes these trained models to make on-the-fly inferences about whether an intervention should be initiated for a specific participant. Both processes utilize the same pre-processing layer.

\textbf{Preprocessing Layer.}\label{just-a-minute: preprocessing_layer} This layer is responsible for preprocessing both spatial and temporal raw data to fit the models' inputs. Timestamp alignment is performed to ensure all sensor readings are synchronized. Data is segmented into fifteen-minute batches, from which statistical features such as median, mean, mode, and variance are extracted. These features are then standardized along the feature axis to guarantee that batches from different time intervals throughout the day have consistent mean values and standard deviations.

\textbf{Offline Process.} Each night, two threads are initiated in parallel to update our models — one for each model. The preprocessing layer is employed to extract features in both threads. The encoder is updated using the previous day's temporal raw data, while the classification head is retrained from scratch using all available feedback on interventions since the beginning of the study, along with the corresponding sensor data (both temporal and spatial). A custom loss function, comprising a classification loss (MSE) and a budget loss, is used for training. The MSE loss aims to identify the ideal intervention timing to maximize user engagement, whereas the budget loss ensures that interventions adhere to the allocated budget constraints. Upon completion of the retraining, both models are saved to the cloud and integrated into the real-time process for subsequent inference.

\textbf{Real-Time Process.} This process infers the appropriate time to intervene, aiming to maximize the likelihood of user acceptance. It is scheduled to run at 5-minute intervals for all active \participants. The process comprises four sequential modules, detailed below.

\begin{enumerate}
    \item Ingestion: The pre-processing layer loads and processes the streaming temporal and spatial sensor data from each \participant.
    \item Encoding: The trained encoder is utilized on the processed temporal data to generate embeddings.
    \item Enrichment: These embeddings are enriched by concatenating them with the following features:
        \begin{itemize}
            \item Features extracted from spatial sensors to capture static changes in the \participant's condition.
            \item Timestamp-related features, such as the day of the week, hour, and relative time of day, are also included. These features offer context to the temporal data and help identify patterns related to the user's daily routine.
            \item Budget-related features, including the remaining time to intervene and the number of interventions left to deliver.
        \end{itemize}
    \item Model: A fully connected neural network serves as the classification head, outputting the likelihood that a proposed intervention will be accepted at that moment. A post-processor then evaluates these odds against the current budget constraints to decide if it's the appropriate time to intervene.
\end{enumerate}

\subsubsection{User Interface (UI) Component}
Recognizing that users often prefer multi-functional apps over specialized ones~\cite{torous2020multiple} and that social media platforms significantly influence app choices~\cite{al2023people}, we have chosen WhatsApp, a widely-used communication app, as the platform for delivering our interventions. This strategy aims to maximize user engagement and minimize the time a \participant needs to adapt to using a new application on a daily routine (up to eliminate it at all). 

\subsection{Stress Intervention Design}
\label{subsection: \mpe Design}

The stress intervention encompasses common coping strategies like \textbf{Relaxation Techniques} and \textbf{Cognitive Strategies}. \textbf{Relaxation techniques}, including deep breathing and mindfulness meditation, have been proven effective in reducing anxiety and stress~\cite{Healthy_Coping_Skills,filaire2007motivation,john2011effect}. Rhythmic activities like walking and yoga are also interventions aiding in stress reduction~\cite{cairney2014uses}.

\textbf{Cognitive Strategies} involve techniques such as cognitive restructuring and goal-setting, aimed at altering stress perceptions~\cite{16_Effective_Stress_Management,Healthy_Coping_Skills}.
The interventions are organized into four categories based on Gross's emotional regulation types~\cite{gross2015emotion} and further classified into four therapy groups: positive psychology, cognitive behavioral, meta-cognitive, and somatic practices~\cite{paredes2014poptherapy}. We developed micro-intervention, quick exercises lasting up to 60 seconds, designed to integrate seamlessly into daily routines. These micro-interventions consist of short videos or text prompts. More details are provided in Appendix~\ref{appendix: examples of interventions and structure}.

\subsection{Study Design}
\label{subsection: Study Design}
We employed a four-week, real-time, in-the-wild design, focusing on parents of young children. During the study, the participants engaged with our WhatsApp chatbot which delivered stress-coping micro-interventions.

\subsubsection{Participants}
We recruited participants through social media, specifically targeting parents of children aged up to 12 years who use \textit{Android} devices, to participate in a four-week study. An initial screening for demographic information and mental health was conducted, leading to the enrollment of 103 participants. Of these, 72 (70\%) provided consent to participate. A total of 37 participants successfully installed the data collection application, while eight encountered compatibility issues, leaving us with a final sample of 29 participants. These individuals were then randomly assigned to one of two conditions, taking into account gender balance due to its known influence on individual stress perception~\cite{american2010stress, american2019stress}. Participants were reimbursed with gift cards of the equivalent of $\sim$15 USD for every phase of the study in which they participated. We refer to phase as a period of two weeks in the study (a total of two phases).

\begin{figure}[tph]
    \centering
    \includegraphics[width=0.65\textwidth]{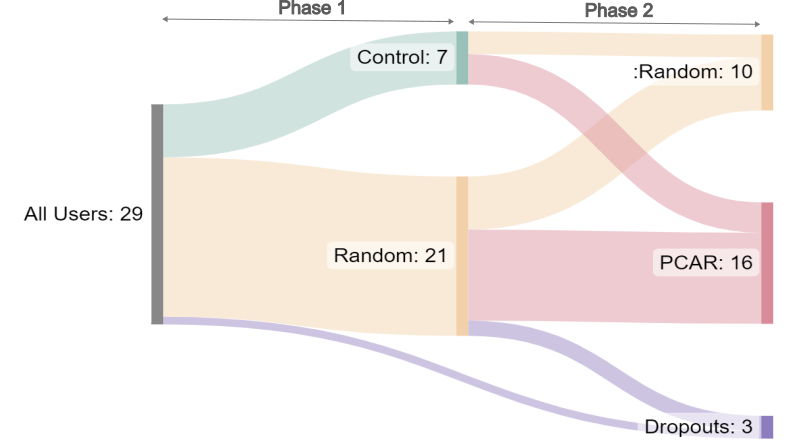}
    \caption{\textbf{Phase-wise User Cluster Differentiation:} Initial allocation included 7 users in the control group and 21 in the random intervention group (Phase 1). After two weeks, 10 users remained in the random intervention group, while 16 were transitioned to \pcar intervention (Phase 2). 1 user didn't complete any phase and 2 users completed the first phase, totalling to 3 dropouts in the study.}
    \label{fig:users_intervene_types.png}
\end{figure}

During the study, one participant didn’t complete any phase and two participants withdrew after the first two weeks (first phase) and did not continue to the subsequent phase. The fluctuations in participant numbers throughout the study are illustrated in Figure~\ref{fig:users_intervene_types.png}.

The demographic breakdown is as follows: 17.2\% male and 82.8\% female; age ranges were 10\% (20-29), 80\% (30-39), and 10\% (40-49); 16\% had one child, 47\% two, 17\% three, and 20\% four or more; 90\% were married or cohabiting; educational levels were 13.3\% high school, 16.7\% undergraduate, and 50\% graduate or higher.

\subsubsection{Procedure and Interactions}
\label{subsection: User Perspective}
\begin{figure}[tph]
    \centering
    \includegraphics[width=0.8\textwidth]{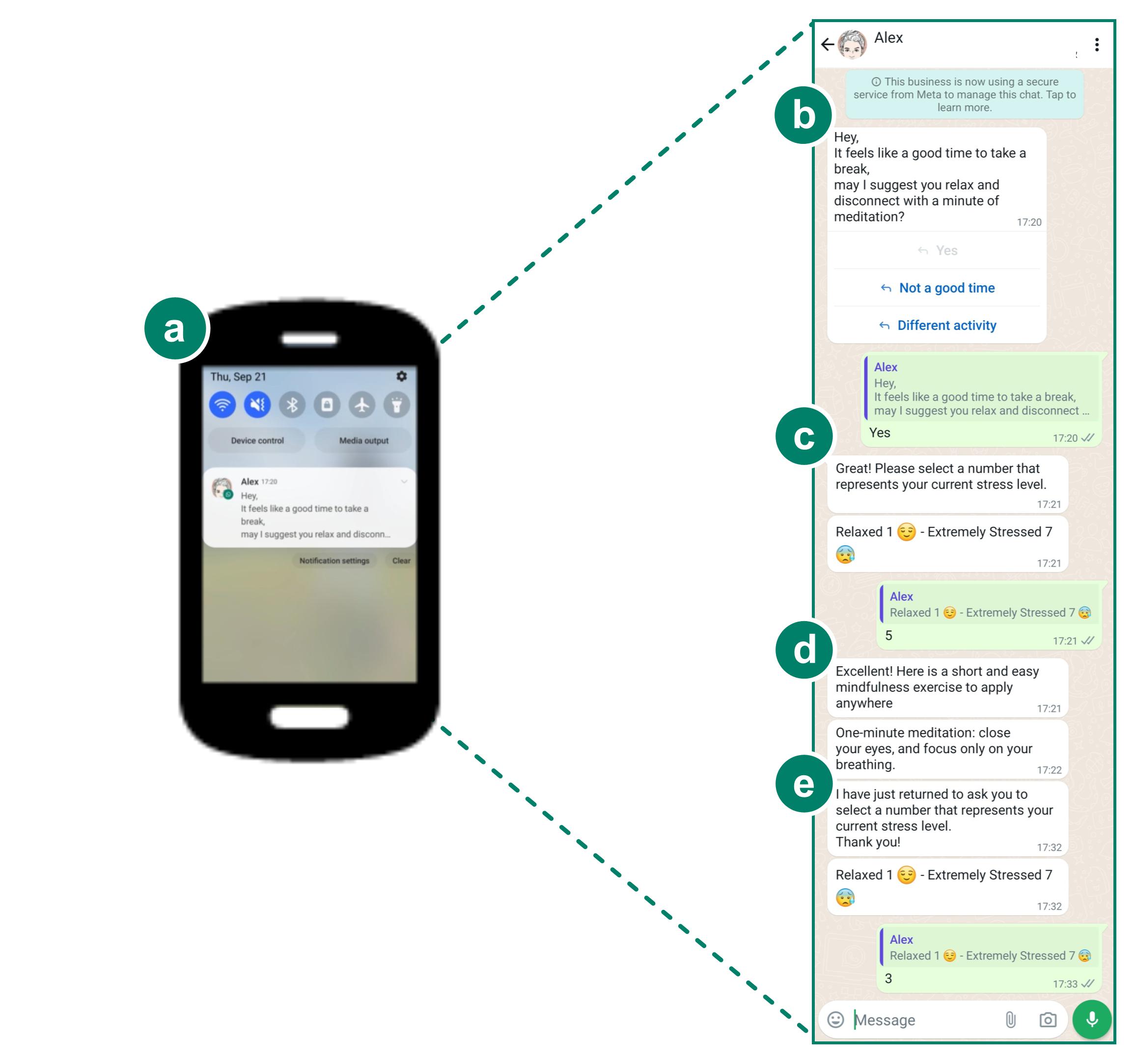}
    \caption{\textbf{Stress Reduction Micro-Intervention Example Dialog:} (a) Our system inferred an ideal time for a stress reduction micro-intervention and initiated a WhatsApp conversation. (b) If the user decides to accept the intervention and engage with our chatbot, he is asked whether it's a good time to intervene. (c) Once the user replies "Yes", he is asked for the current stress level. (d) Our chatbot suggests a stress-coping activity. (e) 10 minutes later, our chatbot asks again for the current stress level.}
    \label{fig:stress_intervention_illustration}
\end{figure}

The study flow was structured as follows:

\begin{itemize}
    \item \textbf{Onboarding} Upon joining the study, participants completed pre-study surveys, including demographic information and psychological scales: the Perceived Stress Scale (PSS)~\cite{cohen1983global} and Big Five Inventory (BFI-10)~\cite{rammstedt2007measuring}. They were guided through the data collection application installation online, running in the background throughout the study only showing a notification "active data collection" (see Section~\ref{subsection: System Design}).

    \item \textbf{Study Phases}
    \label{method:Study_Phases}
    The study was divided into two phases, each lasting two weeks. Interventions were restricted to be delivered on weekdays only from 8AM to 9PM. During the first two weeks (defined as phase 1), participants interacted with our WhatsApp chatbot, which prompted them with either random micro-intervention suggestions or only asked to fill in momentary stress levels if they were assigned to the control group.
    
    After two weeks, the \participants Post-phase PSS questionnaires were administered and participants were divided into two new groups, one receiving personalized context-aware recommendations (\pcarNS) interventions while the other group receiving random interventions (Random group) when the right time to intervene was detected. 
    
    \item \textbf{Feedback and Adaptation} We collected two types of feedback during the study: A binary label on whether the intervention's timing was good or not and pre- and post-intervention stress level on a 7-point Likert scale (1=Not at all stressed; 4=Moderately stressed; 7=Extremely stressed)~\cite{likert1932technique}. Every interaction began with participants receiving a message asking if the current moment was suitable for an intervention. If it was, they were asked to rate their pre-intervention stress level, engage in the suggested activity, and then rate their post-intervention stress level 10 minutes after engaging in a micro-intervention. An example of a full micro-intervention dialog is shown in Figure~\ref{fig:stress_intervention_illustration}.
    Participants also had the option to initiate micro-interventions by messaging the chatbot but this option was never used.
    
    \item \textbf{Offboarding} At the end of the four-week study, participants were guided through the process of uninstalling the data collection application. They then completed a post-study usability questionnaire and a post-experiment PSS questionnaire.
\end{itemize}

\subsection{Data Processing \& Analysis}
\label{subsection: data processing and analysis}
Our analysis was divided into two parts, the first was to understand the engagement of the \participants during the study, and the second was to evaluate the effectiveness of the interventions and their ability to help \participants reduce their short- and long-term stress levels.

Overall, 1207 unique interventions were initiated by our chatbot during the study. A conversation with the chatbot remained active for an hour, that is, if a \participant didn't respond to the chatbot, the session would be closed after an hour and the conversation couldn't be completed, resulting in labeling it as not a good timing for intervention. 
With that policy in mind, 51.2\% (618/1207) of the interventions were initiated, accepted, and engaged to some extent. 91.6\% (566/618) of the accepted interventions were fully completed and collected both pre- and post-intervention momentary stress levels via EMAs. In total, We collected 1179 stress level ratings, 613 pre-intervention and 566 post-intervention.
Study-long stress reduction was calculated using intake, middle (after two weeks), and final PSS scores. All PSS forms during the study were filled appropriately by 26 \participants, totaling with 78 PSS scores measured. The other 3 \participants filled in 1 PSS score at the beginning of the study. 

Momentary stress reduction (\textit{reward}) was computed by subtracting the pre-intervention stress levels from the post-intervention stress levels. A higher value indicates a higher stress reduction. Study-long stress reduction was computed by calculating the gradient of the calculated PSS scores along the study for each \participant. Negative values of the gradient indicate higher stress reduction. 

\subsection{Statistical Tests}
\label{methods: statistical tests}
For comparing the means of two conditions, two examined groups (Control, Random, Personalized), we used the Welch Two Sample t-test. Wherever applicable.
We used one-way analysis of variance (ANOVA) tests to examine differences in outcome variables, such as stress reduction and intervention acceptance rate, considering multi-level factors like intervention activities. If the test p-value was below a significance level, indicating a difference in at least one group mean, a post hoc test, Tukey’s Test in our case, was used for pairwise comparisons while managing the family-wise error rate.
Pearson’s correlation was used for correlation analyses.

We employed the Generalized Estimating Equations (GEE) approach to analyze correlated data with repeated measures, focusing on stress reduction and intervention acceptance rate from the participant perspective. Our GEE model, which accounted for intrinsic correlations, incorporated multi-level attributes like patient day and gender. Due to the time-related nature of our data, we considered correlations to be highest between adjacent time points, with a systematically decreasing correlation as the distance between time points increased. Therefore, we used the first-order autoregressive (AR(1)) working dependence structure as the covariance structure. For the distribution families, the Binomial distribution was used to investigate the acceptance rate, whereas the Gaussian distribution for the stress reduction. The logit function was used as the model's link function. 
 A comprehensive list of features is available in Appendix ~\ref{appendix: gee features description}. This method provided insight into predominant trends while considering individual variations.

All the mentioned procedures for processing the data and for statistical analyses in this study were performed using Python 3.8.8. (statsmodels 0.13.2 was used to establish the GEE models ~\cite{seabold2010statsmodels}).

\section{Results}
\label{section: Finding}
In this section, we present our key findings. The main metrics we follow are the acceptance rate, measuring the willingness of participants to engage when the system triggers a conversation, and the stress reduction, measuring the efficacy of the interventions.
When a metric is displayed as an average across samples, it is calculated as the mean of means, that is, we initially calculated the average for each \participant, followed by an overall average across users. The corresponding standard deviation, in that case, is the standard deviation of the mean. 

\subsection{Impact of Stress Interventions (RQ1)}
\label{finding: Impact of Stress Interventions}
Here, we present data and analysis on the short- and long-term impact of micro-interventions on self-reported stress levels. The results are displayed across different intervention types, comparing no intervention group (control), random interventions, and personalized, context-aware interventions selected by PCAR to evaluate the impact of personalization.

\subsubsection{Short-Term Stress Reduction}
\label{findings: Stress Reduction}
\begin{figure}[tph]
  \includegraphics[width=0.8\textwidth]{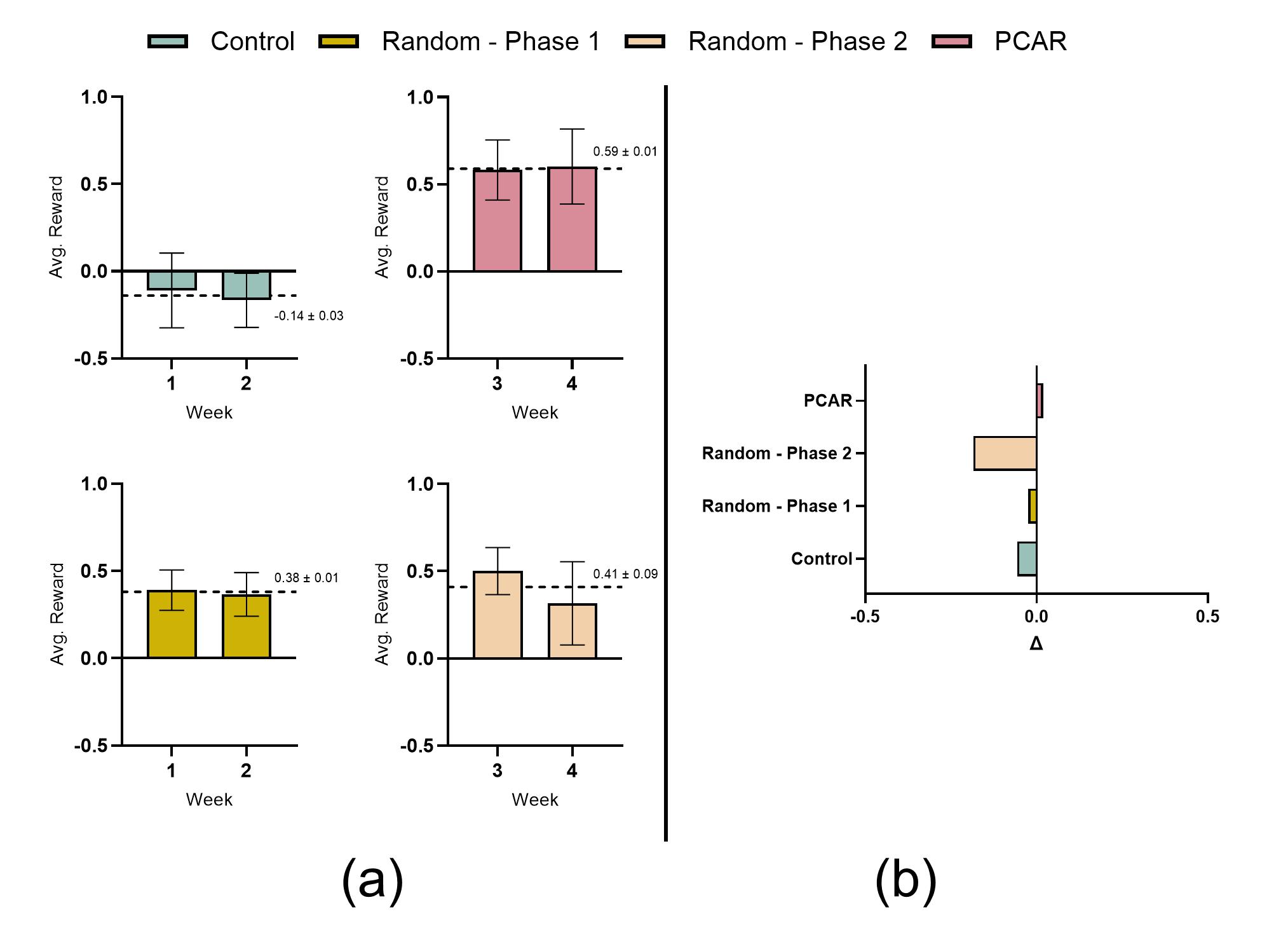}
  \caption{\textbf{Average Weekly Stress Reduction Across Phases and Groups:} (a) This figure is divided into two sections, each corresponding to a different study phase as outlined in Section~\ref{method:Study_Phases}. The y-axis shows the "reward", calculated as the difference between initial and subsequent stress levels on a 7-point Likert scale. The x-axis is divided into weeks. Error bars denote 95\% confidence intervals. We see that for PCAR, the efficacy is maintained week over week while for the other settings, there is a decrease, especially in phase 2. We also observe a significant stress reduction within intervention groups. (b) This figure shows the change in the average reward of each group over the corresponding weeks within the phase.}
  \label{fig:reward_over_time}
\end{figure}

The key finding in this section is that with PCAR, we observe the maintenance of efficacy over time and even a slight increase, as illustrated by the positive $\Delta$ in Figure~\ref{fig:reward_over_time}(b), whereas for the other intervention types, we observe a decrease.

Figure~\ref{fig:reward_over_time}(a) illustrates the average weekly stress levels reported by participants, segmented into the two phases as outlined in Section~\ref{method:Study_Phases}. The term "reward" is used to denote the processed feedback, calculated as the difference between initial and subsequent stress levels. A positive reward indicates a reduction in stress levels, as quantified on a 7-point Likert scale (1=Not at all stressed; 4=Moderately stressed; 7=Extremely stressed). 
We see that the interventions groups had a positive impact on stress reduction in both phases, while the control group exhibited a marginal increase in stress levels. Figure~\ref{fig:reward_over_time}(b) summarizes these changes within the phases to demonstrate the changes over time.

A Generalized Estimating Equations (GEE) model was employed to assess the effects of the random interventions in comparison to a control group during the initial phase of the study (see Section~\ref{methods: statistical tests}). The analysis revealed a statistically significant difference in self-reported stress levels pre- and post-intervention across the two conditions (p=0.001) with a coefficient of 0.32, favoring the random group. This result substantiates the immediate, short-term efficacy of the stress interventions. 

Although existing literature posits that "intervention fatigue" may arise from impersonal or repetitive content, thereby reducing the effectiveness of interventions~\cite{paredes2014poptherapy, laforgue2022last}, our study did not find a significant difference in stress reduction between the random groups across different phases. Similarly, no significant difference was observed between the first phase random group and \pcar group in the second phase.

\begin{figure}[tph]
  \includegraphics[width=0.8\textwidth]{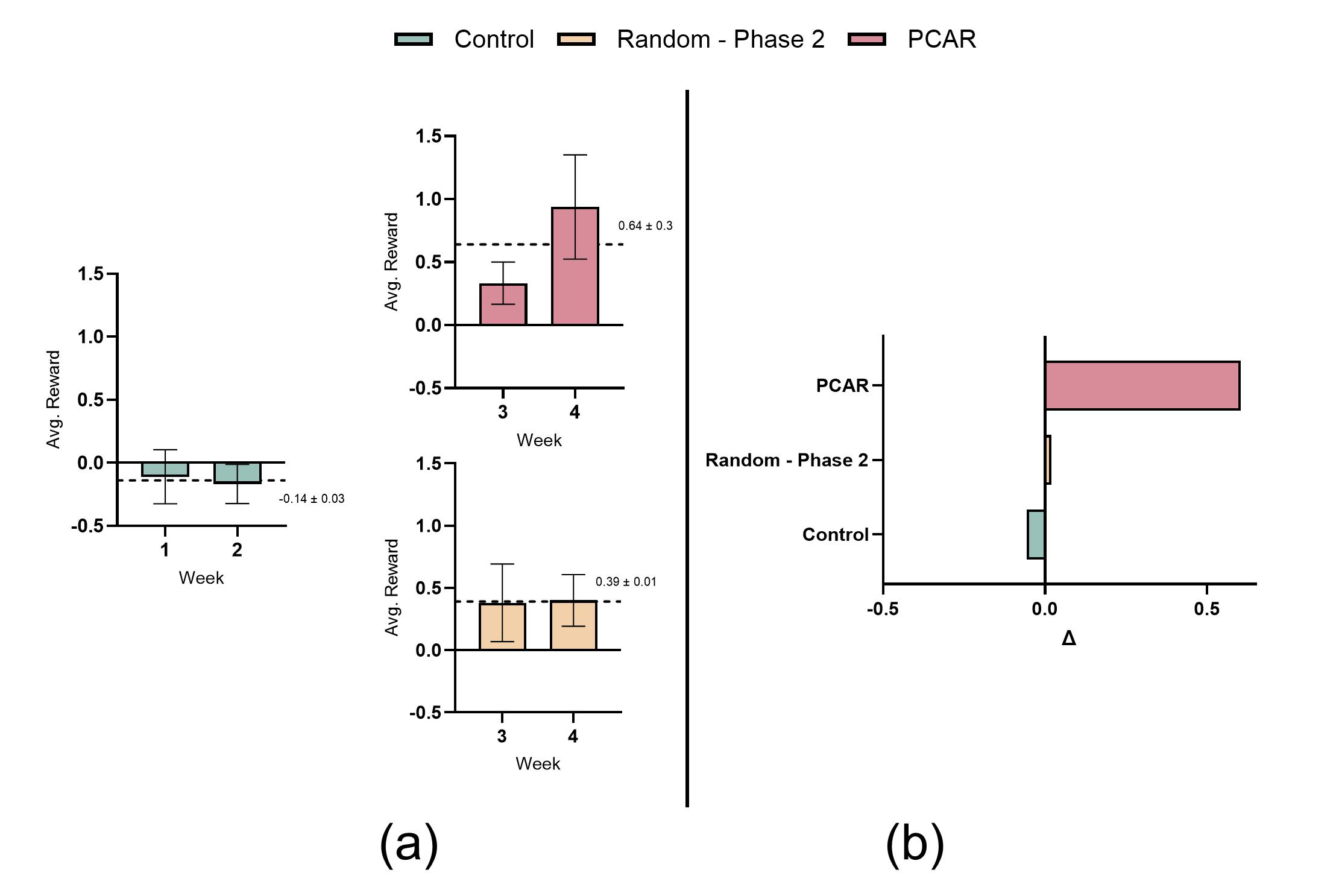}
  \caption{\textbf{Average Weekly Stress Reduction Across Phases of Control Group:} (a) This figure is divided into two sections, each corresponding to a different study phase as outlined in Section~\ref{method:Study_Phases}. The y-axis shows the 'reward,' calculated as the difference between initial and subsequent stress levels on a 7-point Likert scale. The x-axis is divided into weeks. In phase 1, the Control intervention group had a mean reward of -0.14 (\(\sigma = 0.03\%\)), indicating increasing stress. In phase 2, the average reward was 0.39 (\(\sigma=0.01\%\)) for the random interventions group and 0.64 (\(\sigma=0.30\%\)) for the \pcar group, showing stress reduction. Error bars denote 95\% confidence intervals. The graph reveals a significant stress reduction within intervention groups and suggests a preference for \pcar over random interventions. (b) This figure illustrates the change in the average reward for each group over the corresponding weeks within the phase. It underscores the advantage of using personalized, context-aware interventions to enhance participants’ gained value over time.}
  \label{fig:reward_over_time_control_rasa}
\end{figure}

Figure~\ref{fig:reward_over_time_control_rasa} presents a further analysis of the Control group participants ($n=7$). This group of participants was not exposed to interventions during the first phase. The analysis reveals a significant increase in the average reward over time for the PCAR intervention group compared to the random interventions. That is, not only did PCAR succeed in reducing stress and maintaining the positive efficacy of interventions, but its impact was increased over time. We also notice a significant difference ($p=1.1\cdot10^-3$) in the average reward across the two phases.

These results demonstrate the advantage of using PCAR intervention selection algorithm to improve efficacy over time and highlight the significant impact of even a one-minute intervention for stress coping.

\begin{figure}[tph]
  \includegraphics[width=0.8\textwidth]{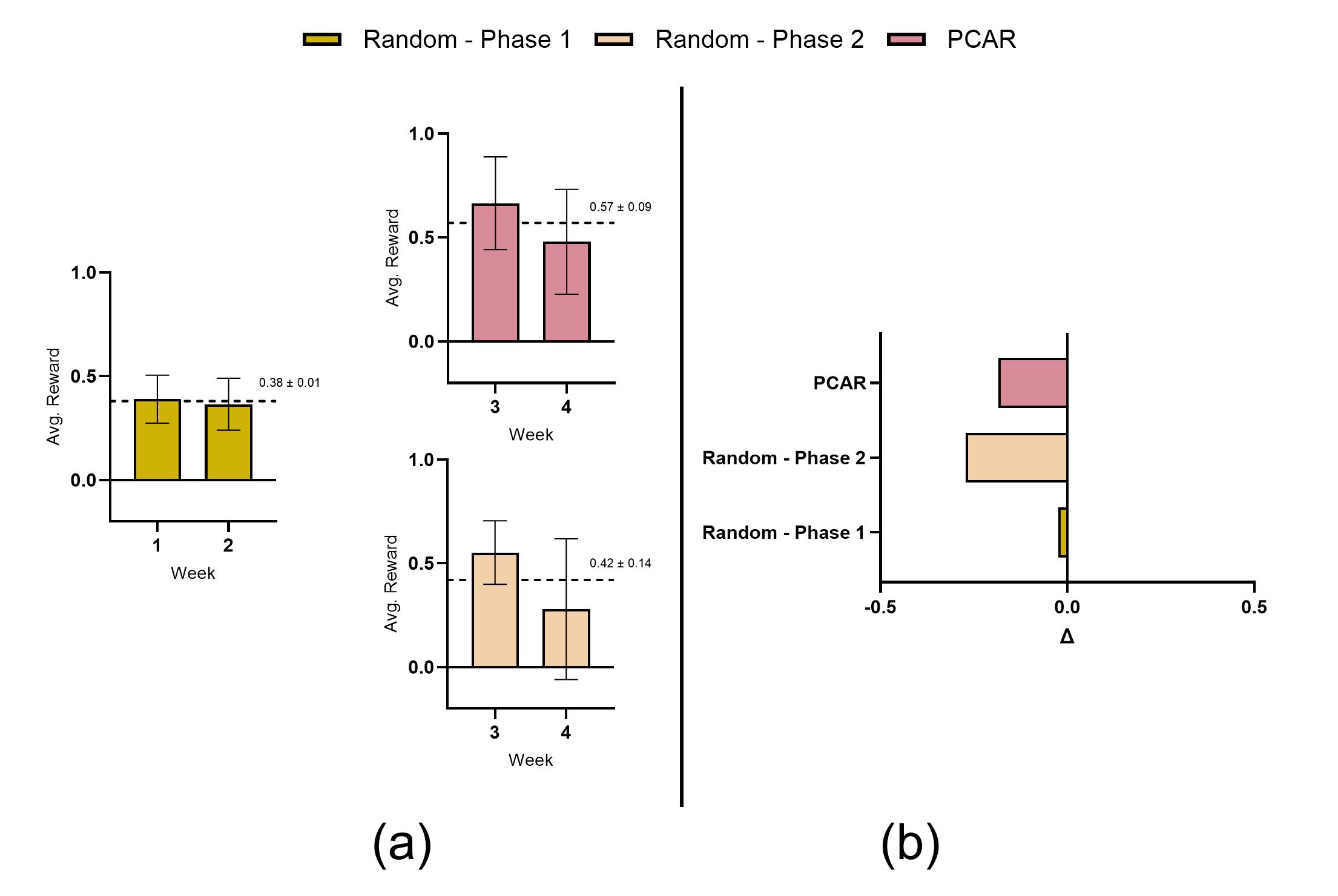}
  \caption{\textbf{Average Weekly Stress Reduction Across Phases of Random Phase 1 Group:} (a) This figure is divided into two sections, each corresponding to a different study phase as outlined in Section~\ref{method:Study_Phases}. The y-axis shows the 'reward,' calculated as the difference between initial and subsequent stress levels on a 7-point Likert scale. The x-axis is divided into weeks. In phase 1, the Random interventions group had a mean reward of 0.38 (\(\sigma = 0.01\%\)), indicating stress reduction. In phase 2, the average reward was 0.42 (\(\sigma=0.14\%\)) for the random interventions group and 0.57 (\(\sigma=0.09\%\)) for the \pcar group, showing stress reduction again. Error bars denote 95\% confidence intervals. (b) This figure shows the change in the average reward of each group over the corresponding weeks within the phase.}
  \label{fig:reward_over_time_random_rasa}
\end{figure}

Figure ~\ref{fig:reward_over_time_random_rasa} shows the average reward across the phases for the participants who were assigned random interventions during Phase~1 (n=21). The average reward of the random intervention groups was positive in both phases 0.38 \((\sigma=0.01)\) and 0.42 \((\sigma=0.14)\) respectively. The average reward for the \pcar interventions was also positive 0.57 \((\sigma=0.09)\).
Despite the participants' feedback on the interventions being positive, we have observed a decrease in the impact of the interventions over time. Nevertheless, the decrease, for the \pcar group, is smaller than for the group of participants who continued to receive random interventions.

\subsubsection{Long-Term Stress Reduction}
\label{findings: Long-Term Stress Reduction}
\begin{figure}[tph]
  \includegraphics[width=0.6\textwidth]{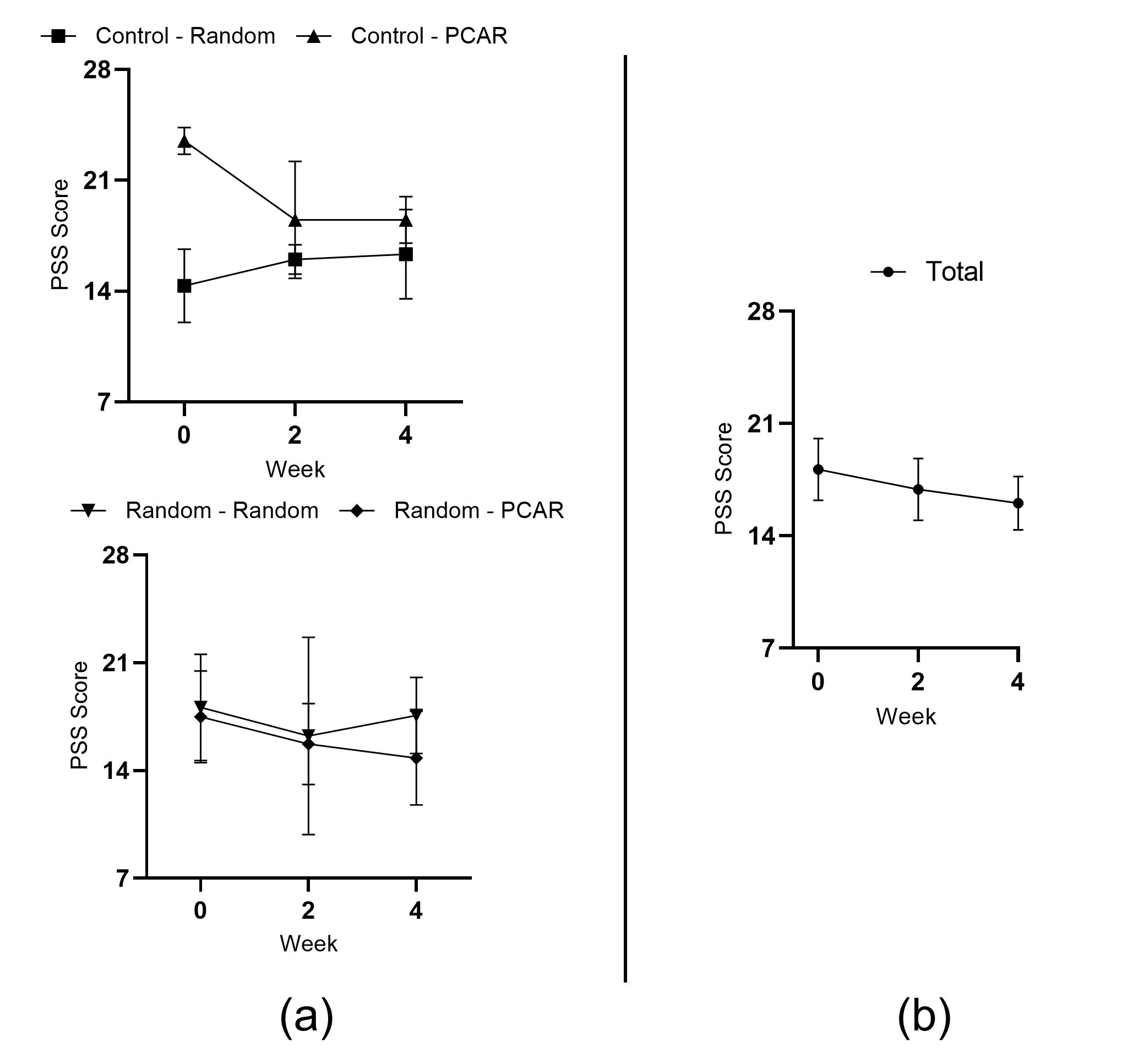}
  \caption{\textbf{Self-reported stress levels throughout the study:} This figure shows the average perceived stress scale from intake, after 2 weeks, and exit surveys. PSS scores of 14-26 would be considered moderate stress. The error bars indicate 95\% confidence intervals. (a) illustrates the average PSS scores across the different study groups. The upper graph represents the participants assigned to the control group in the first phase, while the bottom graph represents those assigned to the random group in the first phase. A decreasing trend is observed after two weeks for participants assigned to the PCAR group, whereas an increasing trend in perceived stress is noted for those receiving more random interventions in the second phase. (b) illustrates The average PSS score of all the participants in the study.
  }
\label{fig:pss_over_time}
\end{figure}

Figure~\ref{fig:pss_over_time}(a) illustrates the average PSS scores across the different study groups. The upper graph represents the participants assigned to the control group in the first phase, while the bottom graph represents those assigned to the random group in the first phase. 

For the bottom graph, a decreasing trend is observed after two weeks for participants assigned to the PCAR group, whereas an increasing trend in perceived stress is noted for those receiving more random interventions in the second phase. in both subgroups the initial average PSS scores and those calculated after the first two weeks exhibit similar values, as both groups received random interventions during this period. This finding highlights the advantage of utilizing PCAR for reducing percieved stress in the long-term.

Pairwise comparisons revealed a significant decrease (\(p \leq 0.0002\)) in perceived stress levels after four weeks of study (PSS scores) with their initial scores. The average PSS score of all the participants dropped from 18.3 to 16.0. PSS scores of 14-26 would be considered moderate stress and below that considered low stress further highlighting the enduring impact of the stress interventions. This decrease, shown in Figure~\ref{fig:pss_over_time}(b), shows the potential for persistent engagement with those interventions in the long term.

\subsection{Timing and Engagement (RQ2)}
In the following section, we will discuss the aspects of timing and engagement in which participants engaged with the proposed micro-interventions.

\subsubsection{Weekly Acceptance Rates Across Phases}
\label{finding: Weekly Acceptance Rate per Phase}
\begin{figure}[tph]
  \includegraphics[width=\textwidth]{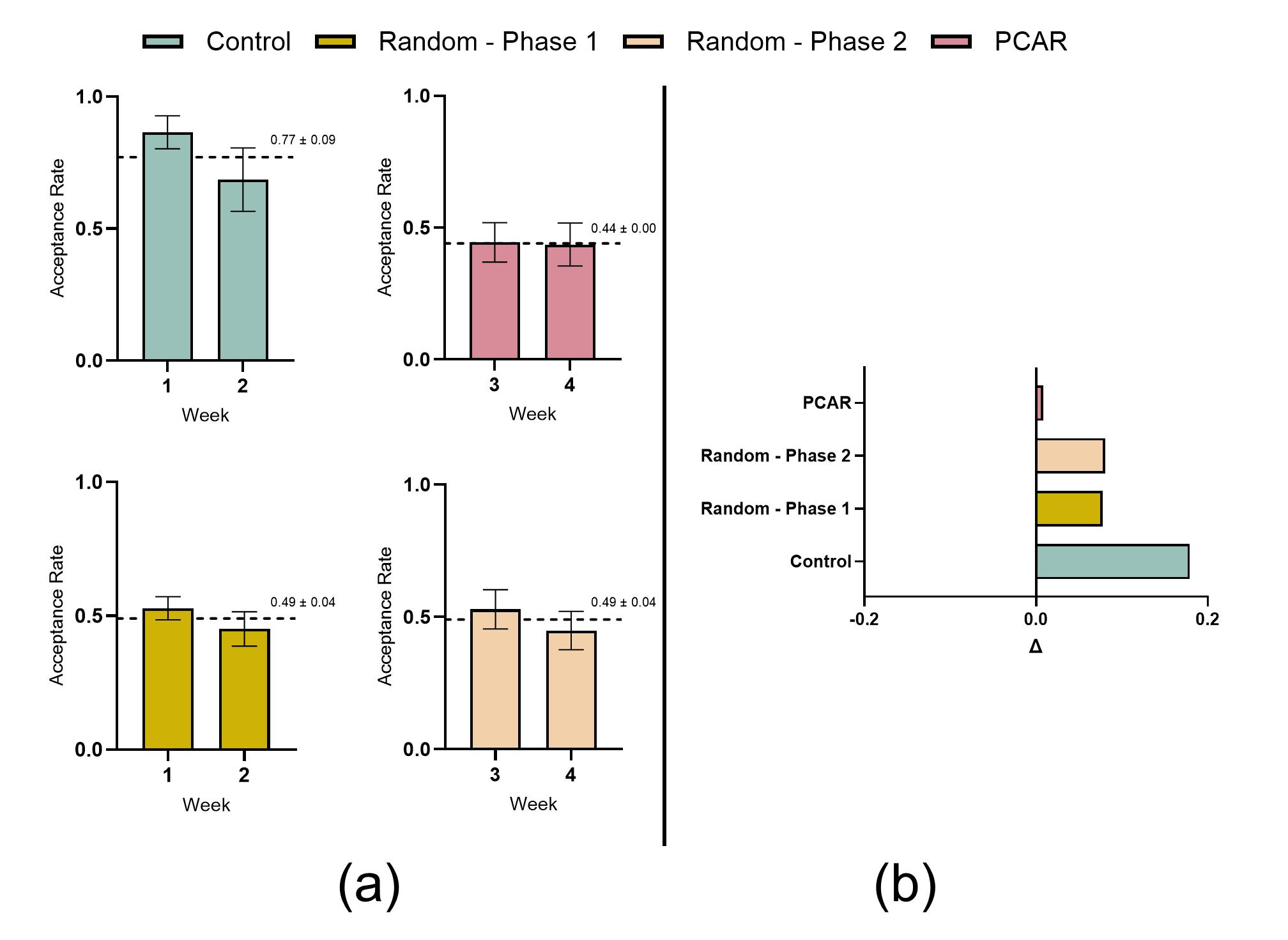}
  \caption{\textbf{Weekly Acceptance Rates Across Phases and Groups:} (a) This figure is divided into two sections, each corresponding to a different study phase as outlined in Section~\ref{method:Study_Phases}. The y-axis shows the acceptance rate calculated as the average number of interventions accepted during the corresponding time frame. The x-axis is divided into weeks. In phase 1, for both the random interventions group and the control group, a decrease in the acceptance rate is observed. In phase 2, the random interventions group's acceptance rate is again decreasing, while the \pcar interventions group's acceptance rate is not affected by the temporal dynamics. Error bars denote 95\% confidence intervals. (b) This figure shows the change in the acceptance rates of each group over the corresponding weeks within the phase. A higher change means a higher retention. The graphs suggests a slight preference for \pcar over random interventions to overcome users retention over time.}
  \label{fig:acceptance_per_phase_all}
\end{figure}

To understand the temporal impact of interventions on participants' retention, we refer to Figure~\ref{fig:acceptance_per_phase_all}. The average acceptance rate for the control group in phase 1 was 0.77 (\(\sigma=0.09\)). The average acceptance rate for the random group in both phase 1 and phase 2 was 0.49 (\(\sigma=0.04\)). The average acceptance rate for the \pcar group in phase 2 was 0.44 (\(\sigma=0.00\)).
The control group achieved the highest acceptance rate among all the tested groups, with a significant difference ($p=1.31e-15$). The \pcar group was the only group that maintained a non-decreasing trend in the acceptance rates over the corresponding weeks within the phase. This observation suggests that personalized interventions are more effective in maintaining engagement with stress-reduction micro-interventions than random interventions.

\subsubsection{Weekly Acceptance Rates (All Participants Altogether)}
\label{finding: Weekly Acceptance Rate}
Maintaining user engagement over time is essential for the effectiveness of a stress management framework. Sustained engagement allows for a more nuanced understanding of user behavior patterns and contributes to the long-term impact of the interventions. Throughout the four-week study period, participants engaged in an average of 22.1 interventions ($\sigma$=2.31\%,  min=4, max=50).

\begin{figure}[tph]
    \centering
    \includegraphics[width=0.6\textwidth]{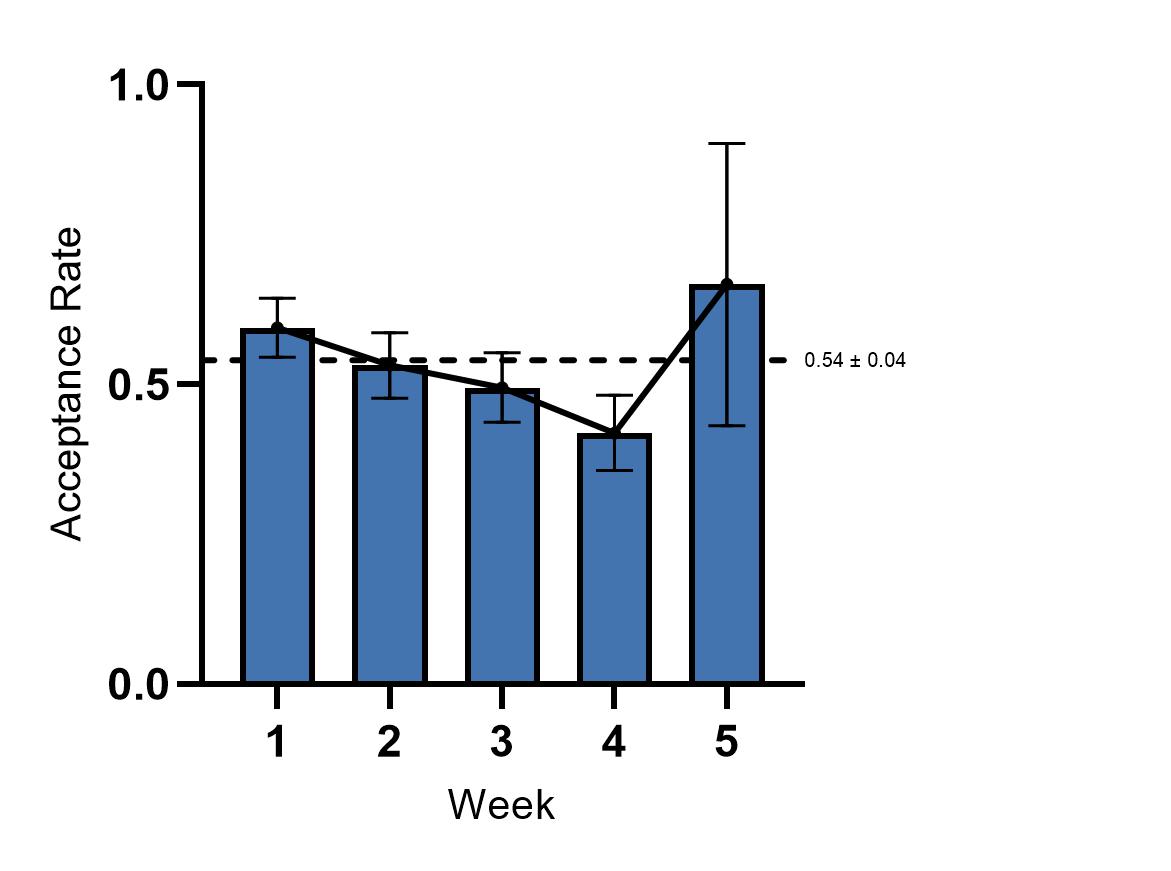}
    \caption{\textbf{Weekly Acceptance Rates of Micro-Interventions:} The figure delineates the trend of acceptance rates over a five-week period, emphasizing a decline in the initial weeks followed by a significant uptick in the fifth week.}
    \label{fig: weekly acceptance rate}
\end{figure}

As depicted in Figure~\ref{fig: weekly acceptance rate}, a noticeable decline in the acceptance rate of micro-interventions is observed across all groups during the main four-week period. Intriguingly, this downward trend is followed by a pronounced increase in the fifth week, which specifically represents participants who voluntarily opted to extend their participation in the experiment beyond the main four-week duration.

We hypothesize that the initial decrease may be symptomatic of "intervention fatigue" or boredom. For the surge observed in the fifth week, we propose that it may be attributed to the specific characteristics of the remaining participants (n=3). These individuals may be more predisposed to accept interventions due to particular personality traits.

This assertion is supported by the observed high variability in participant collaboration. Such variability is manifested in a broad spectrum of acceptance rates with an average rate of 51.6\% $(\sigma=23.7\%, \mbox{min}=0.08, \mbox{max}=1, n=29)$ and engagement levels among different users. Some participants were highly collaborative, consistently accepting interventions and actively engaging with our chatbot throughout the study answering more than 75\% of interventions (n=6). In contrast, others displayed lower levels of collaboration, characterized by intermittent acceptance of interventions and variable engagement levels answering less than 25\% of interventions (n=6). This divergence in behavior could offer an alternative explanation for the sharp increase in acceptance rates observed in the fifth week, a subject we will explore further in the following paragraph.

\subsubsection{Hourly Acceptance Rates}
\label{finding: daily acceptance rate}

\begin{figure}[tph]
    \centering
    \includegraphics[width=0.6\textwidth]{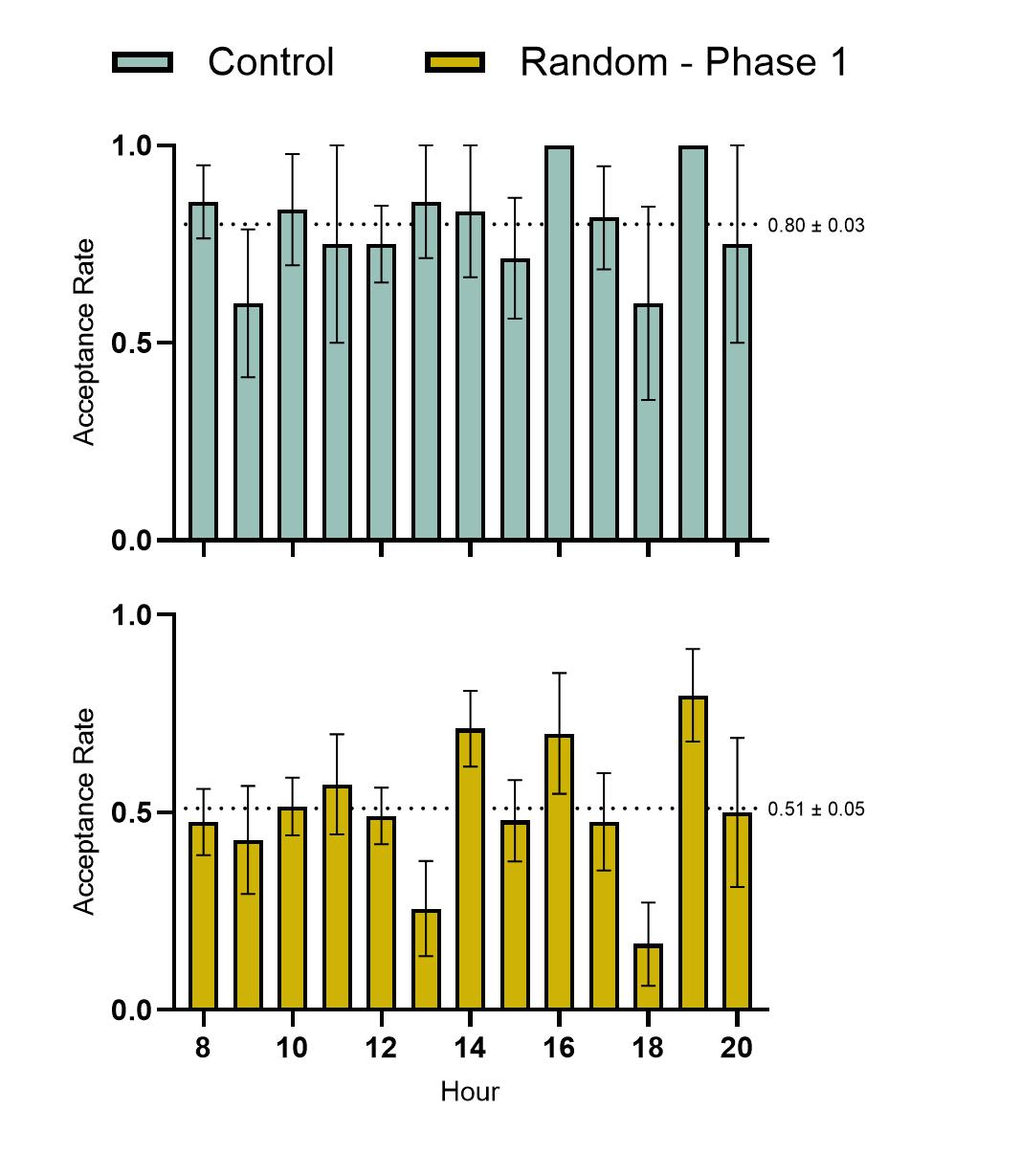}
    \caption{\textbf{Hourly Acceptance Rates of Micro-Interventions:} The figure delineates the average hourly acceptance rates for both control and random groups. Notable peaks and dips correspond with recognized stress-inducing transitional periods for parents of young children.}
    \label{fig: hourly_acceptance_rate.png}
\end{figure}

To elucidate the temporal dynamics of stress and intervention acceptance within and between groups, Figure~\ref{fig: hourly_acceptance_rate.png} presents a histogram that delineates the distribution of intervention opportunities across the day. Each bin signifies the average acceptance rate for a given hour, shedding light on the diurnal patterns of participant engagement with the chatbot.
This histogram is particularly focused on behavior observed during Phase 1. i.e., the first two weeks of the experiment. In this phase, the treatment group— hereafter referred to as the 'random' group — received arbitrary recommendations for engagement, whereas the control group was merely prompted to rate their current stress level. For an in-depth discussion on the study phases, refer to Section~\ref{method:Study_Phases}.

The data reveals dual peaks in acceptance rates at 4 PM and 7 PM for both the control and random groups. Noteworthy declines in acceptance rates manifest at 6 PM. Additionally, the control group experiences a secondary dip at 9 AM, while the random group shows a minor yet pronounced dip at 1 PM. These observed patterns are congruent with extant research on parents of young children, which pinpoints three primary stress-inducing transitional periods: morning, afternoon, and bedtime. Further insights into these periods are elaborated in Section~\ref{subsection: Stress Management in Parents of Young Children}.

While the patterns between the groups are similar, the control group exhibits higher overall acceptance rates, with a rate of 79.7\% ($\sigma$=3.43\%) (n=180 initiated interventions). Conversely, the random group had an acceptance rate of 50.5\% ($\sigma$=4.77\%) (n=480 initiated interventions). The most pronounced divergence between the groups occurred at 1 PM, with the control group showing an acceptance rate of 85.7\% ($\sigma$=14.3\%) compared to 25.6\% ($\sigma$=12.0\%) in the random group. This suggests that even though the intervention takes just a minute, during lunchtime, a minute is highly valuable.

Another significant divergence between the groups occurred at 8 AM. The control group exhibited an acceptance rate of 85.7\% ($\sigma$=9.22\%), in contrast to the random group's rate of 25.6\% ($\sigma$=8.35\%). Given that our system was operational solely on weekdays from 8 AM to 9 PM, this suggests that at 8 AM participants may have already been too busy to engage with an intervention, even for a minute. We hypothesize that earlier interventions could function as mental preparation for the day, potentially leading to higher adherence.

In light of these findings, which demonstrate a significant effect of engagement duration even for one-minute interventions, the optimal timing for interventions should consider two factors: the peak efficacy of the intervention and its alignment with the most receptive moments in the participant's daily routine.

\subsection{Qualitative Assessment (RQ3)}
\label{finding: Qualitative Assessment}
Here We will delve into analyzing the user's personal experience through quantitative and qualitative feedback.
Upon concluding the study, we collected both open and closed feedback from participants. A total of \(n=26\) (89\%) participants contributed to the feedback analysis.

Approximately 61\% of participants expressed satisfaction with the exercises, attributing a positive impact to their daily routine. The content received an average rating of 3.64 on a 5-point scale. Notably, the duration of the exercises garnered the highest mean rating, with nearly half of the participants rating it above 4, indicating strong satisfaction.

Participants generally appreciated the choice of communication platform and found the duration and content of the micro-interventions to be both pleasing and relaxing. These aspects are recommended for retention in future iterations of the study.

On a 5-point agreement scale \((1=\text{Strongly disagree}; 5=\text{Strongly agree})\), participants overwhelmingly agreed that the intervention system facilitated engagement due to an average rating of 4.19 \((\sigma=0.75)\) and found the WhatsApp platform to be convenient with an average rating of 4.29 \((\sigma=0.29)\). The system was not perceived as frustrating, with an average rating of 1.47 \((\sigma=0.23)\).

However, the timing of the interventions received a mixed response, with an average rating of 3 \((\sigma=1.0)\).

No statistically significant differences were observed between groups across all usability metrics.

Open feedback generally reflected satisfaction with the exercises, particularly their utility as mindfulness reminders. A subset of participants (n=4) expressed the expectation for more personalized and timely interventions, highlighting the need for further research in this area.

A point of concern raised in both open and closed feedback was the cumbersome installation process of the \textit{AWARE Light} application(Section~\ref{subsubsection: data collection}). The app's absence from the Google Play Store led to a protracted installation process, which included watching an instructional video and granting multiple permissions. Users also reported a 15\% increase in battery consumption, affecting the overall user experience and resulting in a "good for daily use" an average score of 3 \((\sigma=1.2)\).

\begin{figure}[tph]
  \includegraphics[height=75mm, width=0.8\textwidth]{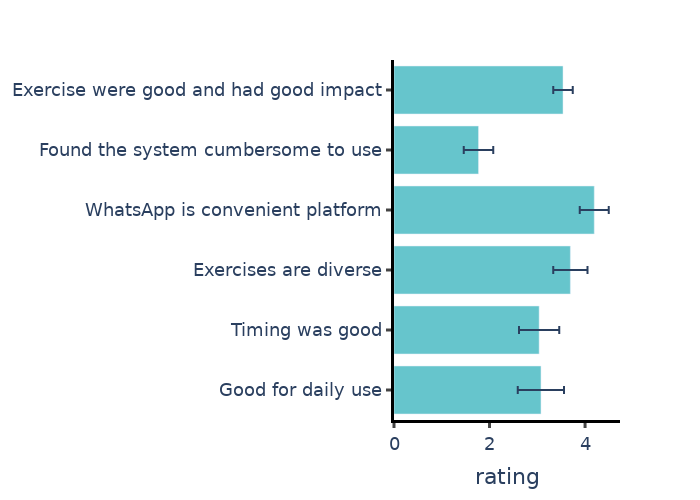}
  \caption{\textbf{Average agreement on six usability statements across n=25 participants, measured on a 5-point Likert scale. The error bars represent 95\% confidence intervals.}}
  \label{fig:Users_review}
\end{figure}

These findings affirm the efficacy of separating the communication platform from the sensing app and underscore the importance of seamless user interaction. The decision to limit micro-interventions to a duration of up to one minute appears to be a significant factor in their positive impact, motivating further research into optimizing timing and content personalization.

\section{Limitations}
\label{section: Limitations}
Conducting our study in a naturalistic setting presented both opportunities and challenges. While this approach enhances the ecological validity of our findings, it also imposes constraints. Specifically, we were limited to intervening only on weekdays, between the hours of 8 a.m. and 9 p.m., and with a maximum of three interventions per day. Additionally, we had limited visibility into other life events affecting our participants. Despite these limitations, we posit that the real-world context of our study lends greater translational value to the results compared to controlled lab experiments.

For a more nuanced understanding and validation of these insights, future research should consider a larger sample size and an extended study duration. Such an approach would provide a more robust assessment of the influence of timing on the effectiveness of stress intervention strategies.

Our study predominantly relied on self-reported measures of stress, which, although valuable, have their own set of limitations. These subjective measures may not always align with physiological indicators of stress~\cite{epel2018more}. They are also susceptible to individual biases, such as reluctance to disclose vulnerability or cultural influences. However, self-reported measures remain a practical choice for in-the-wild studies, as they do not necessitate specialized equipment for data collection and analysis.

Lastly, our data collection was dependent on the AWARE Lite App, which is restricted to Android devices. During the study, we encountered compatibility issues with certain phone models, notably Xiaomi, leading to a significant dropout rate. Of the initial 78 participants who consented to participate and began the installation process, only 30 successfully completed it.


\section{Discussion \& Conclusion}
\label{section: Conclusion}

The prevalence of reported stress has been on the rise, a trend further exacerbated by the Covid-19 pandemic~\cite{american2019stress, jiang2020psychological}. Mobile health applications have the potential to play an important role in coping with the stress crisis due to the ubiquitous use of smartphones which allows the delivery of interventions, particularly micro-interventions~\cite{baumel2020digital}. However, sustaining long-term user engagement with these interventions while preserving their high efficacy remains a challenge. 

In this work we introduce a new algorithm, the Personalized, Context-Aware Recommender (PCAR), for intervention selection to mitigate this issue. We conducted a real-time, field experiment to evaluate PCAR in the wild. Our findings shows that PCAR outperformed random selection in reducing stress and maintaining long-term engagement. 

By investigating the Control group participants, which was not exposed to interventions during the first phase, we noticed that even though the average stress of the Control group participants wasn't reduced in the first phase, once the system integrated micro-interventions within the initiated conversations, the impact was positive and the average stress declined, most noticeably for the PCAR group.

We also observe that while these interventions may require just a minute to complete, there exist specific moments during the day when individuals are more receptive to engaging in them. These moments often coincide with transitions between activities. In our study focusing on parents of young children, we observed higher engagement rates during transitions from work responsibilities to afternoon activities with their children, as well as transitions from afternoon activities to bedtime routines. These insights have implications for the design of interventions, suggesting that they could serve dual purposes: aiding in the "cool-down" process from one activity and acting as mental warm-up exercises for the next engagement. Future work should test intervention delivery timing regarding the stress evolution steps.

\bibliographystyle{ACM-Reference-Format}
\bibliography{references}

\section{Appendix: Experimental Details}
\label{appendix}
\subsection{Raw Sensors Data}
\label{appendix: sensors list}
In the table below, mentioned all the raw sensors data we obtained from \participants' mobile device during the experimental study.

\begin{table}[H]
\centering
\resizebox{\columnwidth}{!}{%
\begin{tabular}{@{}lll@{}}
\toprule
Sensor                          & Data                                                                                                                                  & Type  \\ \midrule
{\color[HTML]{000000} Battery}  & {\color[HTML]{000000} battery information (e.g. power, status)}                                                                       & int   \\
Screen                          & screen status (e.g. turning on, turning off, lock, and   unlock)                                                                      & int   \\
{\color[HTML]{000000} Gravity}  & {\color[HTML]{000000} provides a   three dimensional vector indicating the direction and magnitude of gravity   (x-, y-, and z-axis)} & float \\
Gyroscope                       & Rate or rotation in rad/s around a device (x-, y-, and   z-axis)                                                                      & float \\
{\color[HTML]{000000} Light}    & {\color[HTML]{000000} Level of   ambient light}                                                                                       & int   \\
Linear   Accelerometer          & Acceleration applied to the device, excluding the force of   gravity (x-, y-, and z-axis)                                             & float \\
{\color[HTML]{000000} Location} & {\color[HTML]{000000} location   estimate of the users’ current location}                                                             & float \\
Magnetometer                    & Geomagnetic field strength around the device (x-, y-, and   z-axis)                                                                   & float \\
{\color[HTML]{000000} Rotation} & {\color[HTML]{000000} Orientation of   the device as a combination of an angle and an axis (x-, y-, and z-axis)}                      & float \\
Proximity                       & Distance to an object in front of the device                                                                                          & float \\ \bottomrule
\end{tabular}%
}
\caption{Raw Sensors Description}
\label{tab:raw-sensors-data}
\end{table}

\subsection{GEE Features Description}
\label{appendix: gee features description}
\begin{enumerate}
  \item Daily intervention index - a categorical feature where each sample get the value 1 if the sample was given in the corresponding index along the day.
  \item Intervention counter - counts the number of interventions a user got till the current sample.
  \item Intervention type - a categorical feature denotes the type of intervention a sample got (control, random or sarsa).
  \item Patient day - a number corresponds to the absolute number of days passed since a user joined the study.
  \item Patient week - a number correspond to the number of weeks passed since a user joined the study.
  \item Phase - a Boolean feature equals 1 if the user is in the second phase of the experiment. 
  \item PID (patient id) - user's (cluster) identifier.
  \item Relative time along the day.
  \item Study day - a number corresponds to the absolute number of days passed since the beginning of the study till a sample was given.
  \item Study week - a number corresponds to the absolute number of weeks passed since the beginning of the study till a sample was given.
\end{enumerate}

\subsection{Examples of Interventions and Structure}
\label{appendix: examples of interventions and structure}
Due to the constraints of blind review, a comprehensive list of interventions (n = 116) cannot be disclosed here. However, a link to the full pool of interventions will be made available following the review process.

\begin{figure}[tbh]
    \centering
    \includegraphics[width=\textwidth]{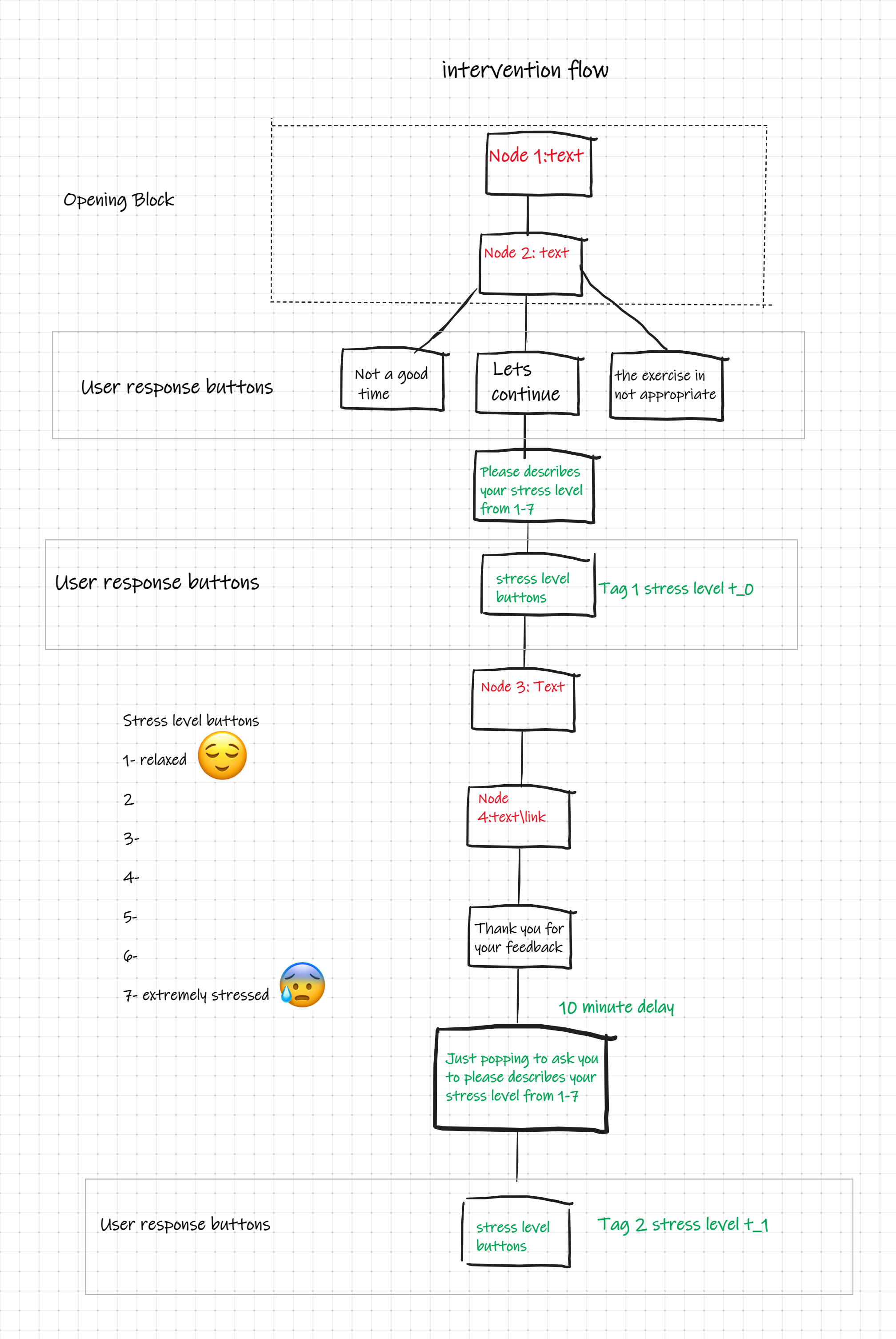}
    \caption{Structure of Stress Intervention, each node represents a configurable text while others remain static.}
    \label{fig: chat_flow.png}
\end{figure}

Presented here sample of interventions:
Each node represents a line in the bot conversation as depicted in Figure~\ref{fig: chat_flow.png}. 

\begin{table}[tbh]
\centering
\resizebox{\columnwidth}{!}{%
\begin{tabular}{@{}llllll@{}}
\toprule
\multicolumn{1}{c}{text}                                                                                                                                                                                         & \multicolumn{1}{c}{node} & \multicolumn{1}{c}{intervention\_type} & \multicolumn{1}{c}{emotional\_regulation} & \multicolumn{1}{c}{therapy\_group} & \multicolumn{1}{c}{location} \\ \midrule
It feels like a good time to take a break                                                                                                                                                                        & node\_id\_1              & meditation                             & response\_modulation                      & meta\_cognitive                    & both                         \\
May I suggest you relax and disconnect with a minute of meditation?                                                                                                                                              & node\_id\_2              & meditation                             & response\_modulation                      & meta\_cognitive                    & both                         \\
Excellent! Here is a short and easy mindfulness exercise to apply anywhere                                                                                                                                       & node\_id\_3              & meditation                             & response\_modulation                      & meta\_cognitive                    & both                         \\
One-minute meditation: close your eyes, and focus only on your breathing.                                                                                                                                        & node\_id\_4              & meditation                             & response\_modulation                      & meta\_cognitive                    & both                         \\
A wise man said that the best time to take a break is when you do not have time for it                                                                                                                           & node\_id\_1              & breathing                              & response\_modulation                      & somatic                            & both                         \\
\begin{tabular}[c]{@{}l@{}}Did you know that even a minute of sports really helps release stress? Sport releases a hormone \\ called endorphin into the body that immediately improves the feeling.\end{tabular} & node\_id\_1              & exercise                               & response\_modulation                      & somatic                            & indoor                       \\
So may I suggest a short stretching exercise? You don't even have to get up from your chair!                                                                                                                     & node\_id\_2              & exercise\_short                        & response\_modulation                      & somatic                            & indoor                       \\
Excellent! Here is a short and easy movement exercise that can be applied anywhere                                                                                                                               & node\_id\_3              & exercise\_short                        & response\_modulation                      & somatic                            & indoor                       \\
Neck stretches: gently tilt your head from side to side, releasing tension in the neck.                                                                                                                          & node\_id\_4              & exercise\_short                        & response\_modulation                      & somatic                            & indoor                       \\
There are moments when distraction is what is most needed.                                                                                                                                                       & node\_id\_1              & scribbling                             & attention\_deployment                     & meta\_cognitive                    & both                         \\
Is now a good time for a fun diversion?                                                                                                                                                                          & node\_id\_2              & scribbling                             & attention\_deployment                     & meta\_cognitive                    & both                         \\
Thanks here is a link                                                                                                                                                                                            & node\_id\_3              & scribbling\_long                       & attention\_deployment                     & meta\_cognitive                    & both                         \\
https://quickdraw.withgoogle.com/                                                                                                                                                                                & node\_id\_4              & scribbling\_long                       & attention\_deployment                     & meta\_cognitive                    & both                         \\
Did you know that music is a powerful healing tool with no side effects and it's also free?                                                                                                                      & node\_id\_1              & music                                  & attention\_deployment                     & meta\_cognitive                    & both                         \\
Can I suggest you a relaxing song?                                                                                                                                                                               & node\_id\_2              & music                                  & attention\_deployment                     & meta\_cognitive                    & both                         \\ \bottomrule
\end{tabular}%
}
\caption{Examples of the types of interventions that participants received during the study}
\label{tab: interventions-examples}
\end{table}

\end{document}